\documentclass[journal]{IEEEtran}
\usepackage{booktabs}

\usepackage{todonotes}
\usepackage{doi}
\usepackage{dblfloatfix}

\usepackage{float}

\usepackage{multirow}
\usepackage{amsmath}
\usepackage{tabularx}  

\usepackage{adjustbox}
\usepackage{amssymb}
\usepackage{amsthm}
\usepackage{natbib}
\usepackage[ruled,vlined]{algorithm2e}
\usepackage{booktabs}
\usepackage{array}
\usepackage{nicefrac}
\usepackage{subcaption}  
\usepackage{setspace}
\usepackage[normalem]{ulem}
\usepackage{caption}
\usepackage{algpseudocode} 
\usepackage{graphicx}
\newtheorem{assumption}{Assumption}
\newtheorem{proposition}{Proposition}
\usepackage[table]{xcolor}
\usepackage{colortbl}
\definecolor{rev}{rgb}{0,0,0}
 
\usepackage{arydshln}
\definecolor{tableheader}{HTML}{D9E1F2}
\definecolor{tablebody}{HTML}{F2F2F2}


\newcommand{\Method}{\textit{CACP}}

\begin{document}

\title{Enhanced Renewable Energy Forecasting\\ using Context-Aware Conformal Prediction}

\author{
\IEEEauthorblockN{Alireza Moradi, Mathieu Tanneau, Reza Zandehshahvar, Pascal Van Hentenryck}\\
\IEEEauthorblockA{
NSF Artificial Intelligence Institute for Advances in Optimization\\
H. Milton Stewart School of Industrial and Systems Engineering\\
Georgia Institute of Technology, Atlanta, GA, USA\\
alirezamoradi@gatech.edu , \{mathieu.tanneau, reza, pascal.vanhentenryck\}@isye.gatech.edu
}
}

\maketitle

\begin{abstract}
Artificial intelligence (AI) is increasingly used to support renewable energy forecasting and grid operations. As renewable penetration grows, reliable probabilistic forecasting is becoming essential for managing uncertainty and supporting risk-aware operational decision-making. However, these forecasts often suffer from miscalibration due to temporal variability, changing weather conditions, and heterogeneous operating regimes. In many real-world settings, renewable energy forecasts are provided by external sources, vendors, or independently trained systems, making retraining infeasible because of limited model access or computational constraints. This creates a need for efficient and model-agnostic methods that can improve forecast reliability after they are produced. This paper presents Context-Aware Conformal Prediction (CACP), a framework for calibrating renewable energy forecasts. The proposed method relies on a weighting mechanism during the calibration procedure which assigns higher weights to historical observations that are more similar to the target forecasting condition. This enables adaptive prediction intervals that reflect local uncertainty regimes without requiring access to, or retraining of, the underlying forecasting model. Experiments are performed on a large-scale dataset from National Renewable Energy Laboratory (NREL) day-ahead solar forecasting, covering multiple systems including MISO, ERCTO, and SPP. The results show that CACP improves the reliability-efficiency tradeoff at both site and system levels compared to NREL's base forecasting model and the other conformal prediction baselines. These results suggest that CACP can serve as a practical reliability-enhancement layer for trustworthy AI-enabled renewable energy forecasting and operational decision support.

\end{abstract}

\begin{IEEEkeywords}
Renewable Energy, Conformal Prediction, Probabilistic Forecasting, Solar Forecasting
\end{IEEEkeywords}

\section{Introduction}
\label{sec:intro}

\subsection{Motivation}


Probabilistic forecasting plays a critical role in modern power systems as operational decisions increasingly rely on accurate predictions of demand, generation, and system conditions. The continued growth of renewable energy generation, especially solar and wind power, has amplified this need as their intrinsic variability introduces substantial uncertainty into grid operations. Recent advances in artificial intelligence (AI) and machine learning have further expanded the use of data-driven forecasting tools for renewable energy, enabling improved prediction of uncertain generation patterns and supporting more informed energy-system operations. In this context, probabilistic forecasting has emerged as an essential tool for planning, market participation, and real-time control in power systems \citep{sweeney2020future,zhang2014review, hong2016probabilistic,hong2020energy,wang2021review}. These methods have been used for short- and long-term forecasting of wind and solar energy \cite{tawn2022review,shi2017direct,golestaneh2016very,xie2023overview}, as well as scenario generation \cite{liao2021windgmmn,zhang2025weather,moradi2025enhanced,dong2022data}.


As AI-driven forecasting tools become central to energy-system operations, the reliability of their uncertainty estimates is increasingly critical. Miscalibrated probabilistic forecasts can result in overconfident or overly conservative decisions, undermining their value for vital downstream applications like reserve allocation, unit commitment, congestion management, and market scheduling. This highlights the dual edge of AI in modern grids: while it could enhance operational decision support, its outputs must be precisely calibrated to manage high-stakes grid operations amid growing renewable uncertainty.



Two key properties of a reliable probabilistic forecasts are: 1) \textit{calibration}, which refers to the statistical consistency between nominal and realized coverage rates, ensuring that predicted probabilities align with observed outcomes; and 2) \textit{efficiency}, which defines the sharpness of forecasts, measured by the tightness of prediction intervals at a given target coverage level \citep{gneiting2007probabilistic,gneiting2014probabilistic}.


Despite the significant advances in statistical \citep{gneiting2023probabilistic,5619586,doubleday2020probabilistic,soman2010review}, physics-based \citep{liu2022use,morales2013renewable,xu2015short}, and deep learning \citep{jonkers2024novel,niu2020wind,xia2021stacked,mashlakov2021assessing} forecasting models, probabilistic forecasts often remain miscalibrated or inefficient, particularly under covariate shifts and regime changes that are common in renewable energy. Moreover, in many operational settings, system operators rely on forecasts generated by external vendors or independently trained models, for which frequent retraining or modification may be infeasible due to limited model access or computational constraints. These challenges call for efficient, model-agnostic approaches that can improve the reliability of probabilistic forecasts after they have been produced.

This paper proposes \emph{Context-Aware Conformal Prediction} (\Method), a new approach based on Conformal Prediction (CP) for calibrating the outputs of probabilistic forecasting models in power systems.
Given a black-box forecasting model, \Method{} generates efficient and calibrated prediction intervals.
The core novelty of \Method{} is a \textit{context-aware weighting} mechanism, utilizing physically relevant features, that assigns higher importance to calibration samples most \textit{similar} to the target conditions at inference time. As a result, CACP can serve as a practical reliability-enhancement layer for trustworthy AI-enabled renewable energy forecasting and operational decision support.


\subsection{Literature Review}
CP emerged as a promising tool for uncertainty quantification and addressing the miscalibration challenge in probabilistic forecasting, offering strong theoretical guarantees and relatively low computational overhead \citep{vovk2005algorithmic}.
Using a hold-out dataset (i.e., calibration set), split CP provides a model-agnostic and distribution-free calibration mechanism that adjusts the prediction intervals to achieve a user-specified target coverage. 
However, conventional split CP relies on the exchangeability assumption \citep{tibshirani2019conformal}, which is not valid in time series.
Due to the non-stationarity and regime changes in time series, particularly in renewable energy forecasting, conventional CP approaches lead to miscoverage or overly conservative prediction intervals \citep{xu2023conformal}.

Recent studies have extended CP to non-exchangeable data and in the presence of covariate shifts. For example, \citep{gibbs2021adaptive} introduces AdaptiveCP, which formulates the distribution shift as a learning problem and dynamically adjusts the target coverage rate to achieve valid coverage. EnbPI is introduced for producing prediction intervals with valid marginal guarantees, utilizing ensemble estimators and bootstrapping \citep{xu2021conformal}. In \citep{xu2023sequential}, the authors introduce SPCI to address the non-exchangeability of time series data and producing efficient prediction intervals by adaptively re-estimating the conditional quantiles of residuals.

Weighting-based methods have been also explored to extend CP beyond exchangeable data by re-weighting the conformity scores in the calibration data. In \citep{tibshirani2019conformal}, the authors propose a weighting mechanism to address covariate shift between the training and test data. NexCP is introduced in \cite{barber2023conformal}, which considers a nonsymmetric algorithm that assigns higher weights to the most recent observations. In \cite{lee2024kernel}, the authors propose KOWCPI which learns data-driven weights for sequential data to enhance the efficiency (i.e., sharpness) while achieving valid marginal coverage. Lastly, neural network-based methods such as HopCPT \citep{auer2023conformal} and CT-SSF \citep{chen2024conformalized} have been recently used for similarity-based sample re-weighting.

In addition to these foundational contributions, recent works have utilized CP for calibration of probabilistic forecasts in power systems. In \cite{jonkers2024novel}, the authors introduced a method combining split CP and quantile random forests for wind power prediction. Similarly, \cite{wang2023conformal} proposes an asymmetric multi-quantile approach for adjusting the prediction intervals in day-ahead wind power prediction. Also, \citep{renkema2024enhancing} implements variants of CP frameworks, weighted CP, CP with KNN, and CP with Mondrian to form uncertainty intervals on top of point-prediction models for solar PV forecasting. These studies differ from this work in that they either rely on specific quantile-based regression models (e.g., linear baselines), were not evaluated on large-scale datasets, or did not include comparisons with comprehensive CP methods.

\subsection{Contributions and Outline}

This paper makes the following key contributions to the advancement of probabilistic forecasting and calibration for renewable energy systems:
\begin{itemize}
    \item A novel context-aware weighted calibration strategy for renewable energy systems, \Method, is introduced. This approach leverages physical characteristics of the system to implement a weighted CP method specifically tailored for renewable energy forecasting.

    \item A dynamic re-calibration framework is introduced, enabling efficient daily adjustment of forecasts to capture dynamic and short-term variations. The approach remains lightweight due to the simple yet effective weighting mechanism of \Method{}, which ensures adaptability without significant computational overhead.


    \item Large-scale experiments are conducted using the National Renewable Energy Laboratory (NREL) day-ahead solar forecasting across multiple U.S. power systems, including \emph{Midcontinent Independent System Operator (MISO)}, \emph{Southwest Power Pool (SPP)}, and the \emph{Electric Reliability Council of Texas (ERCOT)}, at both site and system levels. Namely, at the system level the proposed method improves coverage by approximately $12$--$13$ percentage points (e.g., from $77.5\%$ to $90.1\%$ in ERCOT at the $90\%$ target) while reducing or maintaining interval width. At the site level, where the raw forecasts are substantially more miscalibrated, the gains are larger, improving coverage by up to roughly $42$ percentage points (e.g., from $47.9\%$ to $90.1\%$ in ERCOT).
\end{itemize}

The remainder of the paper is organized as follows. Section~\ref{sec:background} formalizes the problem and reviews the necessary background on conformalized quantile regression. Section~\ref{sec:phycp} introduces the proposed \Method{} framework, including its theoretical motivation, the context-aware weighting strategies and the dynamic recalibration procedure. Section~\ref{sec:ex_setup} describes the experimental setup, covering the NREL dataset, baseline methods, and evaluation metrics. Section~\ref{sec:results} presents the numerical results at both system and site levels, together with conditional-coverage and variable-importance analyses. Finally, Section~\ref{sec:conclusion} concludes and outlines directions for future work.


\section{Problem Definition and Background}
\label{sec:background}

\newcommand{\qhat}{\hat{q}}
\newcommand{\qtil}{\tilde{q}}
\newcommand{\qlhat}{\hat{q}^{\text{l}}}
\newcommand{\quhat}{\hat{q}^{\text{u}}}



    Consider a time series $\{(x_{t}, y_{t})\}_{t \in \mathbb{Z}}$, where  $x_{t} \, {\in} \, \mathbb{R}^{F}$ and $y_{t} \, {\in} \, \mathbb{R}$ denote the \emph{context-vector} and \emph{target variable} at time $t$, respectively.
    The context-vector $x_{t}$ captures relevant contextual information about the target variable (see Section \ref{sec:cacp:features} for a detailed description of the features considered in the paper).
    Let $\mathcal{T}_{cal}$ denote the set of timesteps used for calibration, and let $\mathcal{D}_{cal} \, {=} \, \{(x_{\tau}, y_{\tau})\}_{\tau \in \mathcal{T}_{cal}}$ denote the corresponding calibration set.
    Unless specified otherwise, all time periods are hourly.
    
The paper assumes access to a trained probabilistic forecasting model that outputs predicted quantiles $\qhat^{\alpha}_{t}(x_{t})$, $\alpha \in [0,1]$. This initial model yields prediction intervals of the form
\begin{align}
    \label{eq:background:initial_prediction_interval}
    \hat{C}_{t}^{\alpha}(x_{t})
    =
    \big[
        \qhat^{\nicefrac{\alpha}{2}}(x_{t}),
        \qhat^{1-\nicefrac{\alpha}{2}}(x_{t})
    \big],
    \qquad \alpha \in [0,1].
\end{align}

The goal of the paper is to transform these initial prediction intervals into calibrated prediction sets that satisfy the desired marginal coverage guarantee
\begin{align}
\label{eq: marginal coverage}
    \mathbb{P}
    \big[
        y_{t} \in \tilde{C}^{\alpha}_{t}(x_{t})
    \big]
    \geq
    1 - \alpha,
\end{align}
while maintaining sharp and informative intervals.

To this end, the proposed methodology applies a CP-based calibration procedure, described in Section \ref{sec:phycp}, to adjust the outputs of the underlying forecasting model. This produces calibrated prediction intervals of the form
\begin{align}
    \tilde{C}^{\alpha}_{t}(x_{t})
    =
    \big[
        \qtil^{\nicefrac{\alpha}{2}}(x_{t}),
        \qtil^{1-\nicefrac{\alpha}{2}}(x_{t})
    \big],
\end{align}
where $\qtil$ denotes the adjusted quantiles. The resulting interval $\tilde{C}_t^\alpha$ is designed to satisfy the marginal coverage requirement in \eqref{eq: marginal coverage} while maintaining interval sharpness.

    \emph{\textbf{Remark}}    
    The proposed methodology is agnostic to the underlying probabilistic forecasting model. It only requires access to the model's predictive quantiles, denoted by $\qhat$. These quantiles may be produced directly by a quantile regression model or obtained empirically from a finite set of generated forecast scenarios. Therefore, the calibration procedure can be applied as a post-processing step to a broad class of forecasting models without modifying or retraining them.

\subsection{Conformalized Quantile Regression (CQR)}
\label{sec:background:cqr}

Conformalized Quantile Regression (CQR) 
provides a distribution-free method for calibrating predictive intervals by combining quantile regression with CP \citep{romano2019conformalized}.
A desirable feature of CQR is that it achieves finite-sample coverage guarantees while adapting to heteroscedastic data.

The CQR methodology proceeds as follows.
Let $\alpha \in [0, 1]$, and consider an initial prediction interval $\hat{C}^{\alpha}_{t}(x_{t})$ of the form \eqref{eq:background:initial_prediction_interval}.
Next, for each $\tau \in \mathcal{T}_{cal}$, define the \emph{conformity score}
\begin{align}
    \label{eq:CQR:conformity_score}
    s_{\tau} = \max \bigl(\qhat^{\nicefrac{\alpha}{2}}(x_{\tau}) - y_{\tau},\; y_{\tau} - \qhat^{1-\nicefrac{\alpha}{2}}(x_{\tau})\bigr),
\end{align}
and note that $s_{\tau} \leq 0$ if $y_{\tau} \in \hat{C}^{\alpha}_{\tau}$, and $s_{\tau} > 0$ otherwise.
The conformalized prediction interval is then given by
\begin{align}
    \label{eq:CQR:conformalized_interval}
   \tilde{C}^{\alpha}_{t}(x_{t})
   &= \big[
    \qhat^{\nicefrac{\alpha}{2}}(x_{t}) - \hat{s},
    \qhat^{1-\nicefrac{\alpha}{2}}(x_{t}) + \hat{s}
    \big],
\end{align}
where $\hat{s}$ is the $1{-}\alpha$ quantile of the conformity scores obtained from the calibration data, i.e., 
\begin{align}
    \hat{s} = Q^{1-\alpha}(\{s_{\tau}\}_{\tau \in \mathcal{T}_{cal}}).
\end{align}
The conformalized prediction interval $\tilde{C}^{\alpha}_{t}(x_{t})$ achieves marginal coverage under exchangeability condition \citep{romano2019conformalized}.


While CQR has shown strong performance in many regression settings, its standard form applies a global calibration adjustment and does not explicitly account for temporal dependence, covariate shift, or regime changes. In time-series applications such as renewable energy forecasting, this can lead to over- or under-coverage and unnecessarily wide prediction intervals across different operating conditions. The following section introduces context-aware weighting mechanisms designed to address these limitations by adapting the calibration procedure to the target forecasting context.
\section{Context-Aware Conformal Prediction}
\label{sec:phycp}

\begin{figure*}
    \centering
    \includegraphics[width=0.8\linewidth]{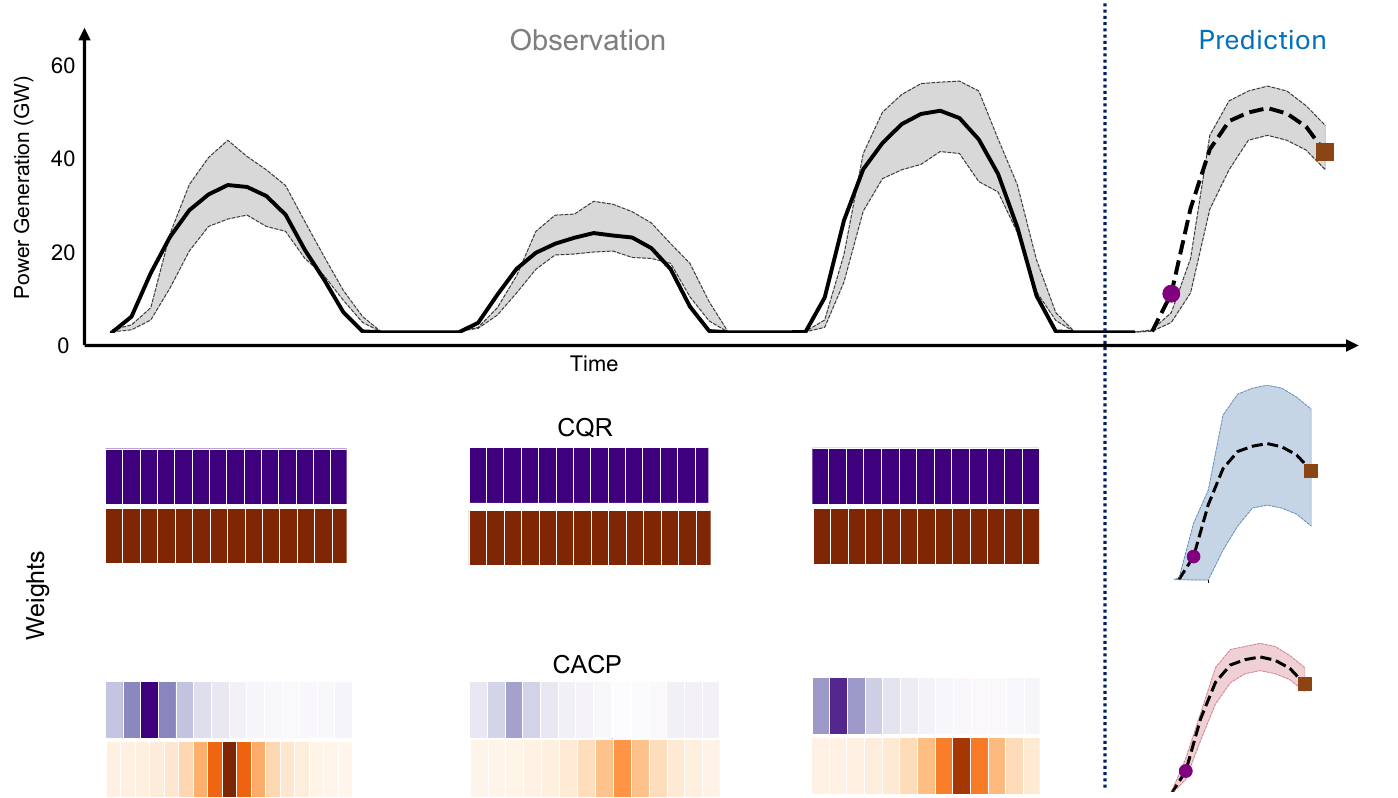}
    \caption{\small Illustration of the proposed CACP framework. Conformity scores from the calibration set are weighted based on their similarity to the target prediction point—illustrated in purple and orange. While CQR assigns uniform weights to all samples, CACP emphasizes more \textit{similar} instances, resulting in tighter and more efficient prediction intervals.}
    \label{fig:cacp}
    \vspace{-1em}
\end{figure*}

Renewable energy systems exhibit strong nonstationarity driven by weather conditions, temporal cycles, and operational variability. Consequently, forecast errors are often regime-dependent, meaning forecasts generated under similar operating conditions tend to exhibit similar conformity score distributions. To account for this structure, \Method{} is introduced as a family of context-aware weighted conformal calibration methods. As illustrated in Figure~\ref{fig:cacp}, unlike CQR, which assigns equal importance to all calibration samples, \Method{} assigns larger weights to calibration samples that are most relevant to the target prediction context. The proposed framework is motivated by weighted CP \citep{barber2023conformal} and recent theoretical developments on CP under temporal dependence \citep{xu2023sequential}. By leveraging physically meaningful auxiliary covariates, \Method{} adapts conformal calibration to local uncertainty regimes in renewable energy forecasting. This section presents the relevant theoretical background on weighted CP and then introduces the proposed method.

\subsection{Theoretical Motivation and Foundations}
\label{sec:theory}


Given a real-valued random variable $Z$ and $\alpha \in [0,1]$, the $\alpha$ quantile of $Z$ is denoted by $Q^{\alpha}(Z)$.
Similarly, let $\{(z_1,\omega_1),\dots,(z_N,\omega_N)\}=\{(z_i,\omega_i)\}_{i=1}^N$ denote pairs of sampled values $z_i$ and corresponding nonnegative sample weights $\omega_i \geq 0$. Define the normalized weights as
\begin{align}
\label{eq:weights}
p_i =
\begin{cases}
\dfrac{1}{\omega_1+\cdots+\omega_{t+1}},
& i=t+1,
\\[8pt]
\dfrac{\omega_i}{\omega_1+\cdots+\omega_{t+1}},
& \text{otherwise},
\end{cases}
\end{align}
which satisfy $\sum_{i=1}^{N} p_i = 1$.
The weighted $\alpha$ quantile is then defined as
\begin{align}
\label{eq:weighted_quantile_def}
Q^{\alpha}\big(\{(z_i,p_i)\}_{i=1}^{n}\big)
=
Q^{\alpha}\left(
\sum_{i=1}^{n} p_i \delta_{z_i}
\right),
\end{align}
where $\delta_x$ denotes the Dirac distribution concentrated at $x$.
Note that, accordingly, $\sum_{i=1}^{n} p_{i} \delta_{z_{i}}$ is a discrete random variable that takes value $z_{i}$ with probability $p_{i}, i \in \{1, ..., N\}$.
When the weights are equal, the quantile is obtained as
\begin{align}
Q^{\alpha}(z_{1}, ..., z_{n}) = Q^{\alpha}(\{(z_{i},\frac{1}{n})\}_{i=1}^{n})    
\end{align}


Now, given a set of calibration instances with nonnegative sample weights $\omega_\tau$ and corresponding conformity scores $s_\tau$, for $\tau \in \mathcal{T}_{\text{cal}}$, the weighted conformal threshold adjustment can be obtained through the weighted quantile construction introduced above. This results in a prediction interval of the form introduced in Eq. \eqref{eq:CQR:conformalized_interval}, where $\hat{s} = Q^{1-\alpha}(\{(s_{\tau}, p_\tau)\}_{\tau \in \mathcal{T}_{cal}})$. In standard CP, all calibration samples contribute equally to the conformal quantile estimation, namely $w_1 = \cdots = w_t = 1$.






Following \citep{barber2023conformal}, the resulting coverage gap is
\begin{align}
\Delta \mathrm{Cov}
=
\alpha - \alpha^\star,
\end{align}
where $\alpha$ is the nominal miscoverage rate and $\alpha^\star$ is the realized miscoverage rate, satisfies
\begin{align}
\Delta \mathrm{Cov}
\geq
-
\sum_{\tau \in \mathcal{T}_{\mathrm{cal}}}
p_{\tau}
\cdot
d_{\mathrm{TV}}(\mathcal{D},\mathcal{D}_{\tau}),
\label{eq:coverage_gap}
\end{align}
where $d_{\mathrm{TV}}(\mathcal{D},\mathcal{D}_{\tau})$ denotes the total variation distance between the original data sequence and the sequence obtained by swapping the test point with calibration point $\tau$.

Equation~\eqref{eq:coverage_gap} provides the main intuition underlying \Method{}. The coverage gap becomes smaller when larger weights are assigned to calibration samples that are distributionally similar to the target prediction point. In renewable energy systems, such similarity naturally emerges from shared operating regimes and meteorological conditions. Consequently, the objective of the context-aware weighting mechanism is not merely to retrieve similar covariates, but rather to identify calibration samples that approximately follow the same forecast error distribution as the target prediction context.

\subsubsection{Operating Regimes and Local Exchangeability}
\label{sec:regimes}
The coverage-gap bound in Eq.~\eqref{eq:coverage_gap} shows that miscoverage is controlled by the weighted total-variation distance between the target point and the calibration samples. This motivates a structural assumption on renewable energy data: that the calibration set decomposes into a small number of recurring \emph{operating regimes}, within which the conformity scores are approximately exchangeable.

Renewable energy systems exhibit such regimes naturally, each associated with distinct physical and meteorological conditions, such as clear-sky midday production, cloudy transition periods, winter daylight patterns, or low-generation evening hours. These regimes induce different uncertainty structures and therefore different conformity score distributions. 
Formally, let $\mathcal{T}_i \subseteq \mathcal{T}_{\text{cal}}$ denote a partition of the calibration time indices $\mathcal{T}_{\mathrm{cal}}$ into latent operating regimes, where each regime $\mathcal{T}_i$ is associated with a conformity score distribution $\mathcal{S}_i$, for $i \in \{1, \dots, R\}$, where $R$ denotes the number of regimes.

\begin{assumption}[Operating Regimes]
\label{ass:regimes}
There exists partitions $\mathcal{T}_i$ of the calibration time steps such that:
\begin{enumerate}
    \item The conformity scores $\{s_{\tau}\}_{\tau \in \mathcal{T}_i}$ are approximately i.i.d.\ within each regime $\mathcal{T}_i$;
    \item Distinct operating regimes induce distinct conformity score distributions:
    \[
        i \neq j
        \quad \Rightarrow \quad
        \mathcal{S}_i \neq \mathcal{S}_j.
    \]
    \item The target prediction point $t$ belongs to a regime $\mathcal{T}_{i}$ identifiable from the context-vector $x_t$;
    
\end{enumerate}
\end{assumption}

In effect, \Method{} replaces unattainable global exchangeability with \emph{local} exchangeability within the target regime.

Following \citep{xu2023sequential}, assuming the within-regime conformity score distribution admits a Lipschitz continuous cumulative distribution function, this local exchangeability guarantees both asymptotic conditional and marginal coverage.

\begin{proposition}[Asymptotic Coverage of \Method{}]
\label{prop:coverage}
Under Assumption~\ref{ass:regimes},
\begin{align}
    \left|
    \mathbb{P}
    \left(
    y_t \in \tilde{C}^{\alpha}_t(x_t)
    \mid x_t
    \right)
    -
    (1-\alpha)
    \right|
    \leq
    g(|\mathcal{T}_{i}|,L_{i}),
\label{eq:coverage_bound}
\end{align}
where
\[
g(|\mathcal{T}_{i}|,L_{i})
\to 0
\quad \text{as} \quad
|\mathcal{T}_{i}| \to \infty,
\]
and $L_{i}$ denotes the Lipschitz constant associated with the conformity score distribution of regime $i$.
\end{proposition}

The proposition follows directly from Theorems~1--2 of \citep{xu2023sequential} applied to the subset of calibration samples retrieved by the context-aware weighting mechanism. The result formalizes the intuition that prediction intervals become increasingly reliable as more calibration data from similar operating regimes becomes available. 

\subsection{Context-Aware Weighted Calibration}
\label{sec:phycp:cacp}

Building on the operating-regime structure of Section~\ref{sec:regimes}, \Method{} operationalizes local exchangeability through a similarity-based weighting of the calibration samples. Let $t$ denote the target prediction time step with context-vector $x_t$. For each calibration sample 
$\tau \in \mathcal{T}_{\mathrm{cal}}$, the method computes a weight
\begin{align}
\label{eq:weight_function}
w_{\tau}
=
\psi(x_t,x_{\tau})
\geq 0,
\quad
\forall \tau \in \mathcal{T}_{\mathrm{cal}},
\end{align}
where $\psi$ is a nonnegative similarity function between the target  context-vector $x_t$ and the calibration context-vector $x_{\tau}$.

Gievn a weighting mechanism $\psi$, the conformalization starts by normalizing the weights $w_{\tau}$ into probabilities $p_{\tau}$ following Eq.~\eqref{eq:weights}. Next, the correction term is obtained
as the weighted quantile of the conformity scores defined in Eq.~\eqref{eq:weighted_quantile_def} as $\hat{s}
= Q^{1-\alpha} \left(\{(s_{\tau},p_{\tau})\}_{\tau \in \mathcal{T}_{\mathrm{cal}}} \right)$. The calibrated prediction interval $\tilde{C}^{\alpha}_t(x_t)$ then follows from the conformalized formulation in Eq.~\eqref{eq:CQR:conformalized_interval}. By assigning larger probabilities to calibration samples within the target operating regime $\mathcal{T}_{\kappa}$, the weighting concentrates the quantile estimate on locally relevant conformity scores, yielding adaptive intervals that tighten the coverage-gap bound of Eq.~\eqref{eq:coverage_gap}.

While any nonnegative similarity function may be employed, the paper considers three computationally efficient and interpretable weighting strategies.

\paragraph{Kernel-Based Weighting}
This approach computes continuous similarity weights using a radial basis function (RBF):
\begin{align}
\psi_{\mathrm{RBF}}(x_t,x_{\tau})
=
\exp
\left(
-\gamma \|x_t-x_{\tau}\|^2
\right),
\end{align}
where $\gamma > 0$ is a hyperparameter controlling the locality of the weighting function.

\paragraph{KMeans-Based Weighting}
This strategy partitions the calibration covariates $\{x_{\tau}\}_{\tau \in \mathcal{T}_{\mathrm{cal}}}$ into $K$ clusters $\mathcal{D}_1,\dots,\mathcal{D}_K$ using K-means clustering. Let $k$ denote the cluster containing the target point $x_t$. The weighting function is then defined as
\begin{align}
\psi_{\mathrm{kmeans}}(x_t,x_{\tau})
=
\begin{cases}
1,
& x_{\tau} \in \mathcal{D}_{k},
\\
0,
& \text{otherwise}.
\end{cases}
\end{align}

\paragraph{KNN-Based Weighting}
Given a neighborhood size $K$, let $\mathcal{D}_K(x_t)$ denote the set of $K$ nearest neighbors of $x_t$ within the calibration set. The similarity weights are then defined as
\begin{align}
\psi_{\mathrm{knn}}(x_t,x_{\tau})
=
\begin{cases}
1,
& x_{\tau} \in \mathcal{D}_K(x_t),
\\
0,
& \text{otherwise}.
\end{cases}
\end{align}

To enable efficient nearest-neighbor retrieval over large calibration sets, the implementation employs a KD-tree data structure, reducing the computational cost of neighborhood queries compared to exhaustive pairwise distance calculations.

\subsection{Context-Aware Features for Solar Generation Forecasting}
\label{sec:cacp:features}

    \Method{} leverages auxiliary features in the covariate vector $x_{t}$, which encode physically-relevant information.
    These include \emph{historical lags} to model short-term dependencies, \emph{time embeddings} to represent periodic temporal patterns, \emph{weather forecasts} to capture atmospheric conditions driving solar generation, and \emph{solarity} to capture daily operational cycles.
    
    \paragraph{Historical Actuals}
    \label{sec:phycp:features:lag}
        This feature encodes recent historical values of the actual generation to capture short-term temporal dependencies. where for an actual $y_t$ the past k observations with a lag value $l$ denoted as
        \begin{align}
            H^l_{t,k} = [y_{t-l},\dots,y_{t-l-k}] .
        \end{align}
        The lag value $l$ ensures that only actual generation values available at the time of calibration are used. For example, when calibrating a forecast in the middle of the day, actual observations from later hours of the same day are not yet available and are therefore excluded.

    \paragraph{Time Embeddings}
    \label{sec:phycp:features:time}

        \newcommand{\thetah}{\theta^{\text{h}}}
        \newcommand{\thetam}{\theta^{\text{m}}}
        \newcommand{\thetad}{\theta^{\text{d}}}

        Solar generation exhibits pronounced temporal patterns, including diurnal and seasonal variations. These patterns are encoded using time embeddings:

        \begin{equation}
        \label{eq:features:temporal}
        \xi^{(v)}_{t} = (\sin \theta^{(v)}_{t}, \ \cos \theta^{(v)}_{t}), \quad 
        \theta^{(v)}_{t} = \frac{2\pi a^{(v)}_{t}}{P^{(v)}},
        \end{equation}
        where $v \in {\text{h}, \text{d}, \text{m}}$ denotes the temporal period (hour, day, or month, respectively), and
        $a^{(\text{h})}_{t} \in \{1, \ldots, 24\}$, $a^{(\text{d})}_{t} \in \{1, \ldots, 365\}$, and $a^{(\text{m})}_{t} \in \{1, \ldots, 12\}$ represent the hour of the day, day of the year, and month of the year, respectively.
        The corresponding periods are $P^{(\text{h})} = 24$, $P^{(\text{d})} = 365$, and $P^{(\text{m})} = 12$. Note that sin-cos embeddings are standard in machine learning literature for embedding temporal information, as they naturally encode the periodicity of the underlying data.

    \paragraph{Normalized Time of Solar Day (Solarity)}
    \label{sec:phycp:features:solarity}

        One limitation of the time embeddings defined in Eq. \eqref{eq:features:temporal} is that they do not explicitly capture variations in daylight throughout the year.
        This is especially important when considering a system comprising multiple solar sites spread across a large geographical area.
        To that end, the paper considers an additional embedding $\xi^{s}_{t} \, {=} \, (\sin \varphi_{t}, \cos \varphi_{t})$, where $\varphi_{t} \, {=} \, 2 \pi \rho_{t}$ and $\rho_{t}$ is the \emph{normalized time of solar day}, defined as:
        \begin{align}
            \rho_{t} &= \frac{t - t^{\text{sunrise}}}{t^{\text{sunset}} - t^{\text{sunrise}}} \in [0, 1],
        \end{align}
        where $t$ denotes the current time step, and $t^{\text{sunrise}}, t^{\text{sunset}}$ denote the sunrise and sunset time at a given location, respectively.
        Note that this feature is only defined during daytime.
        When considering the total generation across the entire system, the definition above is extended by defining $t^{\text{sunrise}}, t^{\text{sunset}}$ as the earliest sunrise and latest sunset time across all solar sites in the system.

    \paragraph{Weather Forecasts}
    \label{sec:phycp:features:weather}
    Solar power generation is strongly driven by atmospheric conditions, making day-ahead forecasts a natural feature for context-aware calibration.
    The following variables are used in this study:
    total cloud cover (TCDC), 2-meter air temperature (TMP), surface pressure (PRES), downward short-wave radiation flux (DSWRF), visible beam downward solar flux (VBDSF), and visible diffuse downward solar flux (VDDSF).
    The last three provide complementary views of surface irradiance: DSWRF captures The total broadband flux, while VBDSF and VDDSF decompose the visible component into its direct and diffuse contributions, respectively.
    The weather feature vector at time $t$ is denoted $\xi^{w}_{t} \in \mathbb{R}^{6}$.
    These variables are obtained from the day-ahead forecasts of the High-Resolution Rapid Refresh (HRRR) model~\citep{dowell2022high}, a convection-allowing numerical weather prediction system operated by NOAA, accessed through the Herbie toolkit~\citep{blaylock2022herbie}. 
   

        The covariate vector $x_t$ (same for $x_\tau$) is then formed by concatenating the features as follows:
        \begin{equation}
            {\color{rev} x_t = (H^l_{t, k}, \xi^h_t, \xi^d_t, \xi^m_t, \xi^s_t, \xi^w_t).}
        \end{equation}
        The extended context-vector thus incorporates both temporal structure and atmospheric conditions, enabling CACP to identify calibration samples that are physically similar to the target prediction context.
        Additional covariates such as {solar-geometry features (e.g., solar angle) can be incorporated to further enhance predictive performance. 
        
        To ensure the intervals remain efficient and well-calibrated, the proposed method includes a hyperparameter tuning and feature selection step as part of training. For each weighting mechanism, this process identifies optimal hyperparameters and the most informative subset of features based on validation performance, see Section \ref{sec:phycp:training}.

\subsection{Dynamic Hypertuning and Calibration Strategy}
\label{sec:phycp:training}
An important contribution of the paper is a dynamic hypertuning strategy that improves adaptability to evolving temporal dynamics and potential distribution shifts in the data.
Thereby, hypertuning, feature selection, and model selection steps are performed periodically as follows.

Let $\delta_{rec}$ denote the recalibration frequency, which is set to daily in this study.
Every $\delta_{rec}$ steps, a tuning set $\mathcal{D}_{tune}$ and validation set $\mathcal{D}_{val}$ are formed using historical data. This is done by splitting a held-out dataset, i.e., data not used to train the underlying probabilistic forecasting model, into two subsets, with $\mathcal{D}_{val}$ containing the more recent observations.


For each \Method{} variant, corresponding to a specific choice of the weight function $\psi$ in Section \ref{sec:phycp:cacp}, hyperparameters and feature subsets are selected based on validation performance. This selection is performed using grid search; however, other approaches, such as random search or standard hyperparameter-tuning tools, could also be used. The proposed framework is particularly suitable for this dynamic tuning procedure because the considered weighting mechanisms are lightweight and simple to evaluate, and can be updated frequently in operational settings. In contrast, deep learning-based alternatives such as HopCPT \citep{auer2023conformal} involve multiple hyperparameters and require substantially more computational resources for training, which can limit their scalability for frequent recalibration across many sites or systems.


\section{Experiment Setup}
\label{sec:ex_setup}

\subsection{Dataset}
\label{sec:exp_setup:data}
\begin{table*}[!t]
\centering
\caption{\small Average Performance across Systems (Day-Ahead Forecasts)}
\label{tab:cp_methods_all_systems}
\resizebox{\textwidth}{!}{%
\scriptsize
\begin{tabular}{llrrrrrrrrrrrr}
    \toprule
    \textbf{ISO} & \textbf{Method}
    & \multicolumn{3}{c}{$(1{-}\alpha)=90\%$}
    & \multicolumn{3}{c}{$(1{-}\alpha)=80\%$}
    & \multicolumn{3}{c}{$(1{-}\alpha)=70\%$}
    & \multicolumn{3}{c}{$(1{-}\alpha)=60\%$} \\
    \cmidrule(lr){3-5}
    \cmidrule(lr){6-8}
    \cmidrule(lr){9-11}
    \cmidrule(lr){12-14}
      &  & PICP $\uparrow$ & AIW $\downarrow$ & WS $\downarrow$
      & PICP $\uparrow$ & AIW $\downarrow$ & WS $\downarrow$
      & PICP $\uparrow$ & AIW $\downarrow$ & WS $\downarrow$
      & PICP $\uparrow$ & AIW $\downarrow$ & WS $\downarrow$ \\
    \midrule

    \multirow{8}{*}{\textbf{ERCOT}}
    & NREL & 77.50 & 0.1933 & 0.3455 & 71.98 & 0.1658 & 0.2531 & 65.21 & 0.1418 & 0.2117 & 58.51 & 0.1193 & 0.1839 \\
    \cmidrule(lr){2-14}
    & CQR & 89.05 & 0.2257 & 0.3240 & 80.16 & 0.1741 & 0.2495 & 71.37 & 0.1457 & 0.2097 & 61.82 & 0.1193 & 0.1823 \\
    & AdaptiveCP & 86.40 & 0.2121 & 0.3307 & 77.85 & 0.1716 & 0.2520 & 69.37 & 0.1446 & 0.2114 & 59.41 & 0.1188 & 0.1837 \\
    & NexCP & 89.59 & 0.2272 & 0.3167 & 78.96 & 0.1739 & 0.2452 & 67.50 & 0.1373 & 0.2073 & 56.51 & 0.1081 & 0.1805 \\
    & HopCPT & 87.73 & 0.2144 & 0.3067 & 78.13 & 0.1718 & 0.2419 & 69.14 & 0.1410 & 0.2085 & 60.33 & 0.1143 & 0.1845 \\
    \cmidrule(lr){2-14}
    & \Method\textsubscript{Kmeans} & \uline{89.98} & \uline{0.2144} & \uline{0.2825} & \uline{79.71} & \uline{0.1630} & \uline{0.2280} & \uline{68.75} & \uline{0.1294} & \uline{0.1957} & \uline{58.15} & \uline{0.1028} & \uline{0.1714} \\
    & \textbf{\Method\textsubscript{KNN}} & \textbf{89.36} & \textbf{0.2063} & \textbf{0.2778} & \textbf{80.17} & \textbf{0.1589} & \textbf{0.2222} & \textbf{69.58} & \textbf{0.1267} & \textbf{0.1892} & \textbf{58.51} & \textbf{0.1009} & \textbf{0.1659} \\
    & \Method\textsubscript{Kernel} & 90.13 & 0.2174 & 0.2741 & 79.67 & 0.1650 & 0.2225 & 68.94 & 0.1295 & 0.1913 & 57.41 & 0.1016 & 0.1684 \\
    \midrule[1.2pt]

    \multirow{8}{*}{\textbf{MISO}}
    & NREL & 82.12 & 0.1261 & 0.1386 & 73.01 & 0.0981 & 0.1115 & 66.05 & 0.0792 & 0.0945 & 59.47 & 0.0644 & 0.0819 \\
    \cmidrule(lr){2-14}
    & CQR & 93.29 & 0.1298 & 0.1386 & 84.40 & 0.1003 & 0.1115 & 75.73 & 0.0805 & 0.0945 & 67.87 & 0.0652 & 0.0820 \\
    & AdaptiveCP & 90.64 & 0.1291 & 0.1387 & 79.87 & 0.0998 & 0.1117 & 71.26 & 0.0800 & 0.0946 & 62.33 & 0.0648 & 0.0820 \\
    & NexCP & 92.91 & 0.1285 & 0.1377 & 83.00 & 0.1000 & 0.1114 & 73.68 & 0.0796 & 0.0947 & 60.87 & 0.0603 & 0.0825 \\
    & HopCPT & 91.60 & 0.1105 & 0.1211 & 78.02 & 0.0827 & 0.1000 & 65.64 & 0.0653 & 0.0877 & 56.05 & 0.0520 & 0.0781 \\
    \cmidrule(lr){2-14}
    & \Method\textsubscript{Kmeans} & 88.67 & 0.0930 & 0.1157 & 78.82 & 0.0688 & 0.0933 & 68.06 & 0.0541 & 0.0806 & 58.63 & 0.0435 & 0.0717 \\
    & \textbf{\Method\textsubscript{KNN}} & \textbf{94.62} & \textbf{0.0978} & \textbf{0.1048} & \textbf{84.53} & \textbf{0.0736} & \textbf{0.0851} & \textbf{73.80} & \textbf{0.0588} & \textbf{0.0747} & \textbf{62.94} & \textbf{0.0476} & \textbf{0.0674} \\
    & \Method\textsubscript{Kernel} & \uline{91.88} & \uline{0.0896} & \uline{0.1042} & \uline{80.89} & \uline{0.0684} & \uline{0.0860} & \uline{70.59} & \uline{0.0556} & \uline{0.0752} & \uline{60.51} & \uline{0.0460} & \uline{0.0680} \\
    \midrule[1.2pt]

    \multirow{8}{*}{\textbf{SPP}}
    & NREL & 80.85 & 0.1389 & 0.1846 & 74.93 & 0.1081 & 0.1428 & 68.34 & 0.0872 & 0.1194 & 60.85 & 0.0708 & 0.1032 \\
    \cmidrule(lr){2-14}
    & CQR & 89.92 & 0.1485 & 0.1815 & 82.11 & 0.1106 & 0.1424 & 74.73 & 0.0884 & 0.1192 & 65.31 & 0.0713 & 0.1030 \\
    & AdaptiveCP & 89.65 & 0.1467 & 0.1800 & 80.47 & 0.1102 & 0.1425 & 71.86 & 0.0881 & 0.1194 & 61.93 & 0.0710 & 0.1032 \\
    & NexCP & 90.86 & 0.1500 & 0.1807 & 82.33 & 0.1094 & 0.1419 & 70.97 & 0.0835 & 0.1190 & 58.51 & 0.0633 & 0.1027 \\
    & HopCPT & 93.14 & 0.1297 & 0.1496 & 83.40 & 0.0969 & 0.1231 & 72.60 & 0.0758 & 0.1069 & 62.19 & 0.0602 & 0.0949 \\
    \cmidrule(lr){2-14}
    & \Method\textsubscript{Kmeans} & 90.34 & 0.1163 & 0.1447 & 79.68 & 0.0863 & 0.1204 & 70.24 & 0.0672 & 0.1046 & 60.24 & 0.0538 & 0.0933 \\
    & \Method\textsubscript{KNN} & \uline{94.28} & \uline{0.1175} & \uline{0.1287} & \uline{85.39} & \uline{0.0892} & \uline{0.1069} & \uline{73.89} & \uline{0.0714} & \uline{0.0949} & \uline{62.71} & \uline{0.0584} & \uline{0.0869} \\
    & \textbf{\Method\textsubscript{Kernel}} & \textbf{90.65} & \textbf{0.1034} & \textbf{0.1231} & \textbf{80.55} & \textbf{0.0811} & \textbf{0.1028} & \textbf{70.00} & \textbf{0.0674} & \textbf{0.0921} & \textbf{60.55} & \textbf{0.0573} & \textbf{0.0842} \\
    \bottomrule
\end{tabular}
} 
\end{table*}

Experiments are conducted on large-scale datasets covering major U.S. power systems, including the \emph{MISO}, \emph{SPP}, and the \emph{ERCOT}. The study targets day-ahead solar power generation forecasts at both site and system levels, provided by the NREL~\citep{NREL_PERFORMDataset}.
The NREL PERFORM dataset is selected for these experiments for two main reasons. First, it provides large-scale, high-resolution probabilistic forecasts spanning multiple power systems with substantially different solar generation patterns, weather regimes, and operational characteristics across a wide geographic region, thereby providing a challenging and realistic benchmark for evaluating calibration robustness under varying conditions. Second, it is widely regarded as an important source for model development and benchmarking in renewable energy, offering probabilistic forecasts in the form of quantiles together with corresponding observations across hundreds of geographically distributed solar sites.
\begin{table*}[!t]
\centering
\caption{\small Average Performance across Sites (Day-Ahead Forecasts)}
\label{tab:cp_methods_all_sites}
\resizebox{\textwidth}{!}{%
\scriptsize
\begin{tabular}{llrrrrrrrrrrrr}
    \toprule
    \textbf{ISO} & \textbf{Method}
    & \multicolumn{3}{c}{$(1{-}\alpha)=90\%$}
    & \multicolumn{3}{c}{$(1{-}\alpha)=80\%$}
    & \multicolumn{3}{c}{$(1{-}\alpha)=70\%$}
    & \multicolumn{3}{c}{$(1{-}\alpha)=60\%$} \\
    \cmidrule(lr){3-5}
    \cmidrule(lr){6-8}
    \cmidrule(lr){9-11}
    \cmidrule(lr){12-14}
      &  & PICP $\uparrow$ & AIW $\downarrow$ & WS $\downarrow$
      & PICP $\uparrow$ & AIW $\downarrow$ & WS $\downarrow$
      & PICP $\uparrow$ & AIW $\downarrow$ & WS $\downarrow$
      & PICP $\uparrow$ & AIW $\downarrow$ & WS $\downarrow$ \\
    \midrule

    \multirow{7}{*}{\textbf{ERCOT}}
    & NREL & 47.90 & 0.2072 & 1.3295 & 39.39 & 0.1634 & 0.8410 & 33.75 & 0.1332 & 0.6469 & 29.02 & 0.1087 & 0.5352 \\
    \cmidrule(lr){2-14}
    & CQR & 90.10 & 0.4799 & 0.7095 & 80.31 & 0.3593 & 0.6054 & 70.22 & 0.2786 & 0.5348 & 60.19 & 0.2160 & 0.4788 \\
    & AdaptiveCP & 72.77 & 0.3360 & 0.9910 & 62.51 & 0.2573 & 0.7134 & 54.65 & 0.2043 & 0.5866 & 47.95 & 0.1626 & 0.5051 \\
    & NexCP & 89.04 & 0.4676 & 0.7183 & 79.34 & 0.3547 & 0.6071 & 69.56 & 0.2785 & 0.5347 & 59.69 & 0.2180 & 0.4783 \\
    \cmidrule(lr){2-14}
    & \Method\textsubscript{Kmeans} & \uline{90.09} & \uline{0.4700} & \uline{0.6876} & \uline{80.58} & \uline{0.3563} & \uline{0.5859} & \uline{70.55} & \uline{0.2808} & \uline{0.5185} & \uline{60.60} & \uline{0.2219} & \uline{0.4662} \\
    & \textbf{\Method\textsubscript{KNN}} & \textbf{89.56} & \textbf{0.4598} & \textbf{0.6822} & \textbf{80.26} & \textbf{0.3535} & \textbf{0.5785} & \textbf{70.45} & \textbf{0.2816} & \textbf{0.5119} & \textbf{60.68} & \textbf{0.2253} & \textbf{0.4605} \\
    & \Method\textsubscript{Kernel} & 89.82 & 0.4621 & 0.6866 & 80.62 & 0.3518 & 0.5816 & 70.85 & 0.2783 & 0.5134 & 60.80 & 0.2209 & 0.4610 \\
    \midrule[1.2pt]

    \multirow{7}{*}{\textbf{MISO}}
    & NREL & 71.23 & 0.2039 & 0.3404 & 61.88 & 0.1609 & 0.2676 & 54.05 & 0.1310 & 0.2286 & 46.87 & 0.1068 & 0.2012 \\
    \cmidrule(lr){2-14}
    & CQR & 91.16 & 0.2505 & 0.3070 & 81.52 & 0.1930 & 0.2545 & 71.39 & 0.1543 & 0.2220 & 61.13 & 0.1237 & 0.1977 \\
    & AdaptiveCP & 84.85 & 0.2346 & 0.3121 & 73.82 & 0.1814 & 0.2589 & 64.61 & 0.1458 & 0.2253 & 56.01 & 0.1177 & 0.1998 \\
    & NexCP & 89.64 & 0.2441 & 0.3048 & 79.75 & 0.1893 & 0.2537 & 69.84 & 0.1518 & 0.2217 & 59.94 & 0.1217 & 0.1975 \\
    \cmidrule(lr){2-14}
    & \Method\textsubscript{Kmeans} & 89.60 & 0.2365 & 0.3002 & 79.59 & 0.1820 & 0.2493 & 69.58 & 0.1456 & 0.2179 & 59.58 & 0.1167 & 0.1945 \\
    & \Method\textsubscript{KNN} & \uline{91.26} & \uline{0.2318} & \uline{0.2891} & \uline{81.85} & \uline{0.1784} & \uline{0.2389} & \uline{72.13} & \uline{0.1435} & \uline{0.2090} & \uline{61.90} & \uline{0.1158} & \uline{0.1873} \\
    & \textbf{\Method\textsubscript{Kernel}} & \textbf{90.40} & \textbf{0.2259} & \textbf{0.2886} & \textbf{80.86} & \textbf{0.1742} & \textbf{0.2379} & \textbf{70.93} & \textbf{0.1406} & \textbf{0.2079} & \textbf{60.80} & \textbf{0.1146} & \textbf{0.1864} \\
    \midrule[1.2pt]

    \multirow{7}{*}{\textbf{SPP}}
    & NREL & 58.76 & 0.2095 & 0.8135 & 49.85 & 0.1653 & 0.5422 & 42.76 & 0.1346 & 0.4297 & 36.54 & 0.1098 & 0.3625 \\
    \cmidrule(lr){2-14}
    & CQR & 89.28 & 0.3647 & 0.4940 & 79.14 & 0.2771 & 0.4211 & 69.15 & 0.2177 & 0.3701 & 59.28 & 0.1729 & 0.3303 \\
    & AdaptiveCP & 79.62 & 0.3028 & 0.5808 & 68.33 & 0.2301 & 0.4611 & 58.80 & 0.1820 & 0.3924 & 50.06 & 0.1454 & 0.3440 \\
    & NexCP & 89.41 & 0.3633 & 0.4635 & 79.61 & 0.2851 & 0.4033 & 69.70 & 0.2270 & 0.3597 & 60.01 & 0.1807 & 0.3237 \\
    \cmidrule(lr){2-14}
    & \Method\textsubscript{Kmeans} & 89.30 & 0.3479 & 0.4534 & 79.27 & 0.2731 & 0.3914 & 69.30 & 0.2199 & 0.3493 & 59.41 & 0.1776 & 0.3161 \\
    & \Method\textsubscript{KNN} & \uline{91.70} & \uline{0.3359} & \uline{0.4029} & \uline{82.61} & \uline{0.2693} & \uline{0.3462} & \uline{72.71} & \uline{0.2233} & \uline{0.3116} & \uline{61.77} & \uline{0.1856} & \uline{0.2868} \\
    & \textbf{\Method\textsubscript{Kernel}} & \textbf{90.43} & \textbf{0.3127} & \textbf{0.3880} & \textbf{80.95} & \textbf{0.2524} & \textbf{0.3309} & \textbf{70.86} & \textbf{0.2135} & \textbf{0.2982} & \textbf{60.63} & \textbf{0.1838} & \textbf{0.2757} \\
    \bottomrule
\end{tabular}
} 
\end{table*}
The dataset contains hourly time series of actual and forecast (in the form of quantiles) solar power generation for the year 2019, consisting of a total of 1149 sites (751 for MISO, 172 for SPP, and 226 for ERCOT) and 3 systems. The calibration models take as input NREL's quantile-based forecasts, actual historical values, and context-aware features, and produce adjusted prediction intervals.
For the weather features, each site uses the HRRR forecast at its nearest grid point. At the system level, site-level weather forecasts are aggregated using plant capacities as weights, yielding system-level weather variables that reflect the meteorological conditions experienced by the overall generation fleet. Weather covariates were not available for ERCOT, and the corresponding CACP variants therefore rely only on the historical, time, and solarity features for that system.
The first two months of 2019 (up to March 1st) are used as the initial calibration set. Thereafter, the calibration models are updated daily to adjust the day-ahead forecasts. For each test day, the preceding week serves as the validation set (for parameter tuning), while calibration is performed using all data available prior to that day.



\begin{table}[!t]
\centering
\caption{\small Method hyper-parameters and their values}
\label{tab:param}
\small
\begin{tabular}{lll}
\toprule
\textbf{Method} & \textbf{Parameter} & \textbf{Value} \\
\midrule
CQR & -- & -- \\  
\midrule
AdaptiveCP & $\gamma$ & 1e-4, 5e-4, 1e-3 \\
\midrule
NexCP & $\rho$ & 0.95, 0.98, 0.995 \\
\midrule
HopCPT & $\beta$ & 5.0, 10.0, 20.0 \\
\midrule
\Method\textsubscript{Kmeans} & \textit{K} & 3, 5, 8, 12 \\
\midrule
\Method\textsubscript{KNN} & \textit{K} & 50, 100, 200, 500, 1000 \\
\midrule
\multirow{2}{*}{\Method\textsubscript{Kernel}} 
  & kernel & RBF, Laplacian \\
  & $\gamma$ & 0.5, 1.0, 2.0 \\
\bottomrule
\end{tabular}
\end{table}

\subsection{Baseline Methods}
\label{sec:exp_setup:baselines}

The results are compared with the following baselines:
\begin{itemize}
    \item \textbf{NREL}: The raw solar power generation forecasts provided by NREL. These forecasts are provided in the form of 99 quantiles. For each target coverage rate, the corresponding quantiles are selected. 
    \item \textbf{CQR} \citep{romano2019conformalized}: Applies the CQR method using NREL’s quantile-based forecasts and historical observations.

    \item \textbf{AdaptiveCP} \citep{gibbs2021adaptive}: This approach adjusts the target coverage rate dynamically to account for the distribution drift in the data. The original method is developed for deterministic forecasts. Here, the conformity score is adjusted using \eqref{eq:CQR:conformity_score}.

    \item \textbf{NexCP} \citep{barber2023conformal}: This is a weighting-based CP approach, which considers higher importance to the most recent observation in the calibration data. Similar to AdaptiveCP, the conformity score in NexCP is modified to align with \eqref{eq:CQR:conformity_score}.
    \item \textbf{HopCPT} \citep{auer2023conformal}: This approach trains a neural network to learn similarity-based weights for time series data. The Hopfield network loss function is reformulated for compatibility with quantile forecasts. Due to its high computational cost, HopCPT is evaluated only for system-level forecasts. For this approach, the loss function is defined as below:

    \begin{align}
    \mathcal{L}
    = \frac{1}{|\mathcal{T}|}
    \big\|
    \big(
    |\mathbf{s}_{1:\mathcal{T}}|
    -
    \mathbf{A}_{\mathcal{T}}|\mathbf{s}_{1:\mathcal{T}}|
    \big)^{2}
    \big\|_{1},
    \label{eq:hop_loss}
    \end{align}
where $\mathbf{A}\in \mathbb{R}^{\mathcal{T} \times \mathcal{T}}$ denotes the Hopfield association matrix capturing pairwise dependencies among conformity scores $\mathbf{s}$ as defined in \eqref{eq:CQR:conformity_score}.

    \end{itemize}

Table \ref{tab:param} summarizes the hyperparameters used for all methods. To balance coverage and efficiency, parameters are tuned by minimizing the \textbf{Winkler Score} (WS) as defined in \eqref{eq:ws}, with the exception of HopCPT, whose hyperparameters are optimized by directly minimizing \eqref{eq:hop_loss} over the validation data.

\subsection{Evaluation Metrics}
\label{sec:exp_setup:metrics}
The performance of each method is evaluated using three standard metrics. Let $\mathcal{T}_{\text{test}}$ denote the set of all time indices in the test set, with size $|\mathcal{T}_{\text{test}}|$. Given the target value $y_t$ (for some $t \in \mathcal{T}_{\text{test}}$) and the predicted interval $\hat{C}_t^{\alpha} = [\hat{C}_t^{\alpha,l}, \hat{C}_t^{\alpha,u}]$, the metrics are defined as follows:

\paragraph{Prediction Interval Coverage Probability (PICP)} This metric measures the empirical coverage rate of the prediction intervals and is defined as follows:
\begin{align}
    \label{eq:picp}
    \text{PICP}^{\alpha} = \frac{1}{|\mathcal{T}_{\text{test}}|}\sum_{t \in \mathcal{T}_{\textbf{test}}} \mathbf{1}\{y_t \in \hat{C}_t^{\alpha}\},
\end{align}

\paragraph{Average Interval Width (AIW)} This metric evaluates the sharpness of the prediction intervals at a given target coverage level, $\alpha$, and is defined as: 
\begin{align}
    \text{AIW}^{\alpha} = \frac{1}{|\mathcal{T}_{\text{test}}|} \sum_{t \in \mathcal{T}_{\textbf{test}}} \text{IW}^{\alpha}_t, \\
    \text{IW}^{\alpha}_{t} = \hat{C}_t^{\alpha,u} - \hat{C}_t^{\alpha,l} 
    \label{eq:aiw}
\end{align}

\paragraph{Winkler Score (WS)} This metric combines PICP and AIW and is defined as:

\begin{equation}
    \text{WS}^{\alpha} = \frac{1}{|\mathcal{T}_{\text{test}}|}\sum_{t \in \mathcal{T}_{\text{test}}} \text{WS}_t^{\alpha}
\end{equation}
where,
\begin{align}
    \text{WS}_t^{\alpha} =
    \begin{cases}
        IW_t^\alpha+\tfrac{2}{\alpha}(y_t-\hat{C}_t^{\alpha,u}), & y_t > \hat{C}_t^{\alpha,u}, \\[6pt]
        IW_t^\alpha+\tfrac{2}{\alpha}(\hat{C}_t^{\alpha,l}-y_t), & y_t < \hat{C}_t^{\alpha,l}, \\[6pt]
        IW_t^\alpha, & \text{otherwise}.
    \end{cases}
    \label{eq:ws}
\end{align}
For site-level experiments, all metrics are computed per site and then averaged across all sites that are within the same system. All time series are normalized by the maximum capacity of their corresponding systems or sites.

\section{Numerical Results}
\label{sec:results}

\newcommand{\nrel}{}

\subsection{Results Analysis}


Tables \ref{tab:cp_methods_all_systems} and \ref{tab:cp_methods_all_sites} present the system- and site-level results, respectively. The average PICP$^\alpha$, AIW$^\alpha$, and WS$^\alpha$ values are provided for the proposed method and baselines for MISO, ERCOT, and SPP. Various target coverages are considered, ranges from $60\%$ to $90\%$ (i.e., $\alpha = 0.1,0.2,0.3,0.4)$. All metrics are computed over daylight hours, corresponding to periods with nonzero solar generation. Note that HopCPT is evaluated only for system-level calibration due to the prohibitive computational cost of training across a large number of sites.

The results indicate that the NREL probabilistic forecasts exhibit miscoverage rates ranging from approximately {\color{rev}$9.2\%$ to $22.5\%$} at the system level, and from {\color{rev}$8.3\%$ to $52.1\%$} at the site level {\color{rev}at 90\% target coverage} (see Tables \ref{tab:cp_methods_all_sites} and \ref{tab:cp_methods_all_systems}). The substantially lower coverage at the site level highlights the increased difficulty of forecasting at a high granularity level, due to higher local variability. All CP methods substantially improve the PICP, achieving empirical coverage close to the target across all coverage levels, while maintaining competitive interval widths. This includes the basic CQR approach, which does not include any covariates or context-aware weighting mechanisms. The only exception is AdaptiveCP, which improves upon NREL’s raw forecasts but fails to consistently reach the target, particularly when the base forecasts are severely miscalibrated. For example, in ERCOT, NREL’s PICP {\color{rev}for 90\% target coverage} is 77.5\% at the system level and 47.9\% at the site level. After calibration with AdaptiveCP, the miscoverage remains at 3.6\% and 27.23\%, respectively.

Comparison with baseline methods shows that the proposed $\Method$-based methods consistently achieve the lowest WS for both system- and site-level forecasts, with $\Method_{\text{KNN}}$ being the best performing model. {\color{rev}At the system level, $\Method_{\text{KNN}}$ achieves the lowest WS in 8 out of the 12 system-level settings (three ISOs and four coverage levels), while remaining highly competitive in the remaining cases. At the site level, $\Method_{\text{KNN}}$ attains the best WS for ERCOT and MISO across most coverage levels, whereas $\Method_{\text{Kernel}}$ performs best on SPP. Overall, the proposed context-aware methods consistently outperform CQR, AdaptiveCP, NexCP, and HopCPT in terms of the reliability--efficiency trade-off.} To further illustrate the trade-off between PICP and AIW, Fig.~\ref{fig:picp_aiw_all} plots PICP versus AIW at various target coverage levels for {\color{rev}all} system-level forecasts. As shown, the proposed method consistently outperforms all baselines, achieving tighter intervals while maintaining valid coverage. {\color{rev}Compared with HopCPT, the proposed methods typically achieve similar or better coverage with substantially narrower intervals. For example, in MISO at 90\% coverage, $\Method_{\text{KNN}}$ achieves a WS of 0.1048 compared with 0.1211 for HopCPT, while simultaneously providing higher coverage (94.62\% versus 91.60\%).} Similar experiments were conducted for site-level forecasts, which showed consistent trends.

\subsection{Conditional Coverage Analysis}
{\color{rev}
\begin{figure*}[!t]
    \centering
    \includegraphics[width=\linewidth]{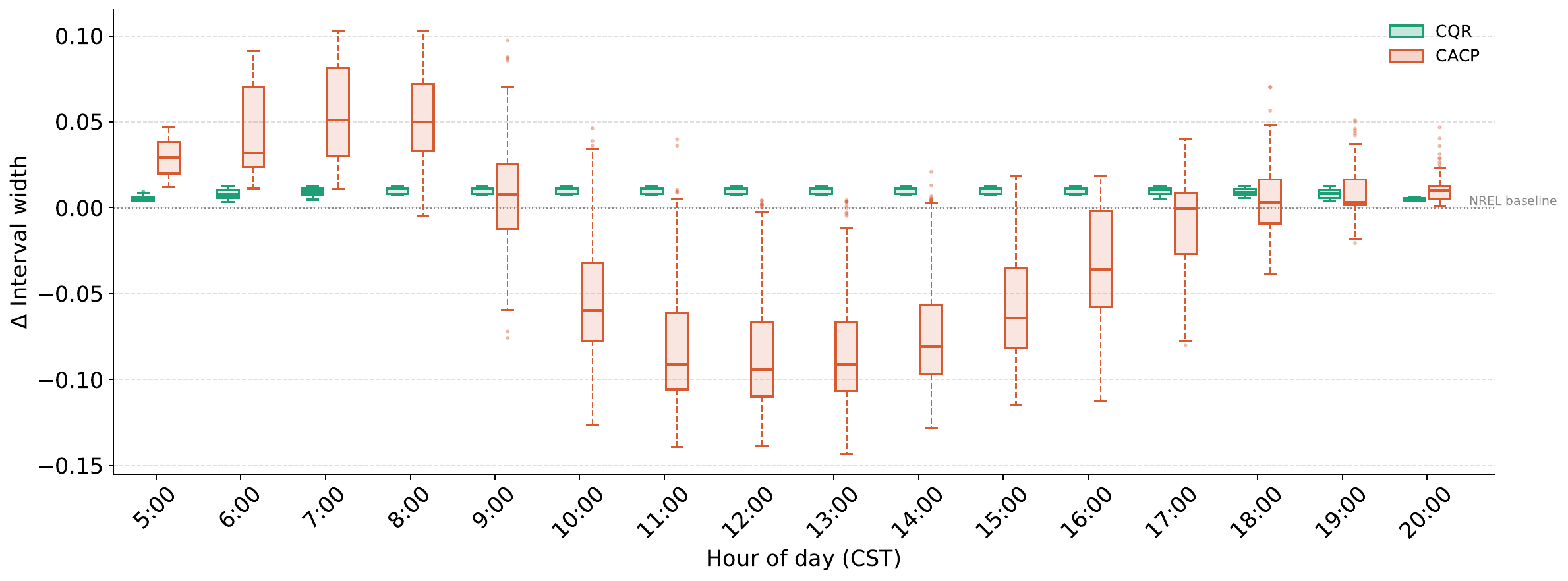}
    \caption{Hourly interval-width adjustment $\Delta_{\text{IW}}$ for the SPP system-level forecasts relative to the original NREL intervals. While CQR applies a nearly constant positive adjustment across all hours, \Method{} exhibits strongly time-dependent behavior, increasing interval widths during high-uncertainty morning and evening periods while reducing interval widths near midday peak-generation hours.}
    \label{fig:delta_hourly}
\end{figure*}
}

This subsection compares the conditional coverage of the proposed method with baseline approaches across different hours of the day. {\color{rev} Fig.~\ref{fig:hourly_picp_all} reports the hourly PICP at the 80\% target coverage level across all system-level datasets.} As illustrated, the proposed method achieves coverage close to the target across all hours, with slight overcoverage during peak hours and mild undercoverage during non-peak periods. In contrast, baseline methods exhibit substantial miscoverage during early morning and evening hours and provide over-confident intervals during the peak hours. These results highlight the superior adaptability of $\Method$ to distributional shifts throughout the day. Similar trends are observed for site-level forecasts, but are omitted here due to space constraints.

To further illustrate this adaptive behavior, Fig.~\ref{fig:delta_hourly} presents the hourly interval-width adjustment $\Delta_{\text{IW}}$ for the SPP system-level forecasts relative to the original NREL intervals for CQR and CACP. As shown, CQR applies a nearly constant positive adjustment across all hours, increasing the width of the initial prediction intervals regardless of operating conditions. In contrast, $\Method$ exhibits strongly time-dependent behavior. During early morning and evening periods, the method increases interval widths to compensate for higher uncertainty and undercoverage. Conversely, during midday peak-generation hours, where the original NREL intervals are already relatively wide, $\Method$ frequently applies negative adjustments, effectively shrinking the prediction intervals while maintaining valid coverage. This behavior demonstrates that the proposed context-aware weighting mechanism adaptively calibrates uncertainty according to the local operating regime rather than relying on a globally fixed correction.

\begin{figure*}
    \centering
    \includegraphics[width=0.5\linewidth]{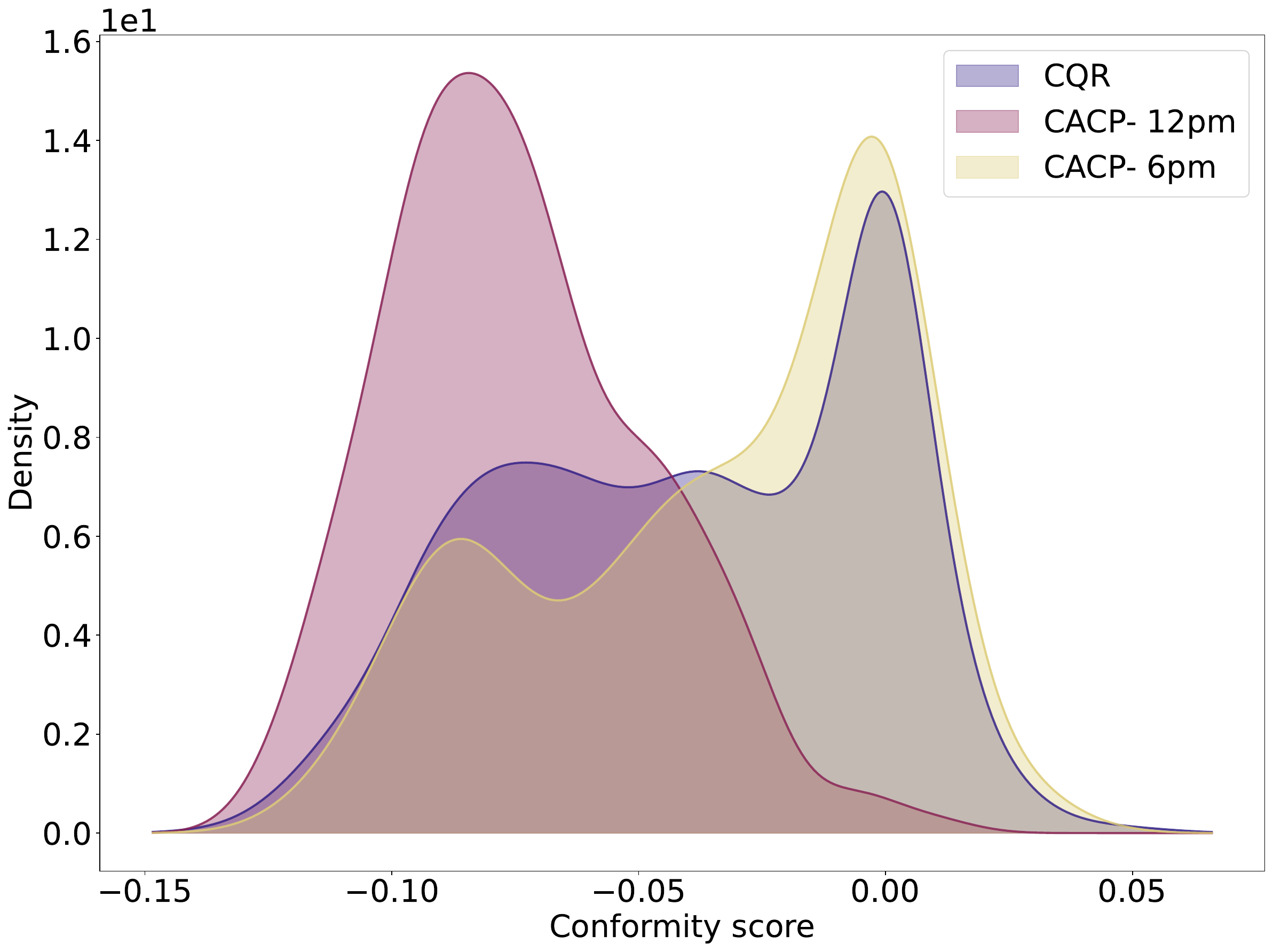}
    \caption{\small Distribution of conformity scores for the \textbf{MISO system-level} forecasts using CQR and CACP at two representative time steps (12pm and 6pm) on 2019-04-2. While the CQR distribution remains relatively stable across hours, the CACP distribution varies with time, showing a right-skewed (more positive) pattern in the evening (6pm) similar to CQR, 
    and a left-shifted (more negative) distribution at noon (12pm).}
    \label{fig:cs_dist}
    \vspace{-1em}
\end{figure*}

To further explain the source of this adaptive behavior, Fig.~\ref{fig:cs_dist} shows the distribution of conformity scores for CQR and $\Method_{\text{KNN}}$ during calibration at two representative hours---6pm and 12 p.m.---on 2019-04-20. Since CQR applies no weighting to conformity scores, the distribution remains unchanged across hours, leading to a fixed small positive adjustment to the prediction intervals (i.e., a uniform increase in width). In contrast, $\Method$ uses context-aware weighting, resulting in distinct conformity score distributions for different hours. At 6 p.m., a non-peak hour, the weighted conformity scores are concentrated near zero, prompting a small positive adjustment and a slight increase in interval width. At 12 p.m., a peak hour when the initial forecast intervals are already wide, the weighted conformity scores are skewed toward negative values, leading to a reduction in interval width. Comparison between the $\Delta_{\text{IW}}$ adjustments for CQR and CACP is for SPP system-level forecasts across different hours of the day is presented in Fig. \ref{fig:delta_hourly}.

\begin{figure*}[!t]
    \centering

    \begin{minipage}[t]{0.32\linewidth}
        \centering
        \includegraphics[width=\linewidth]{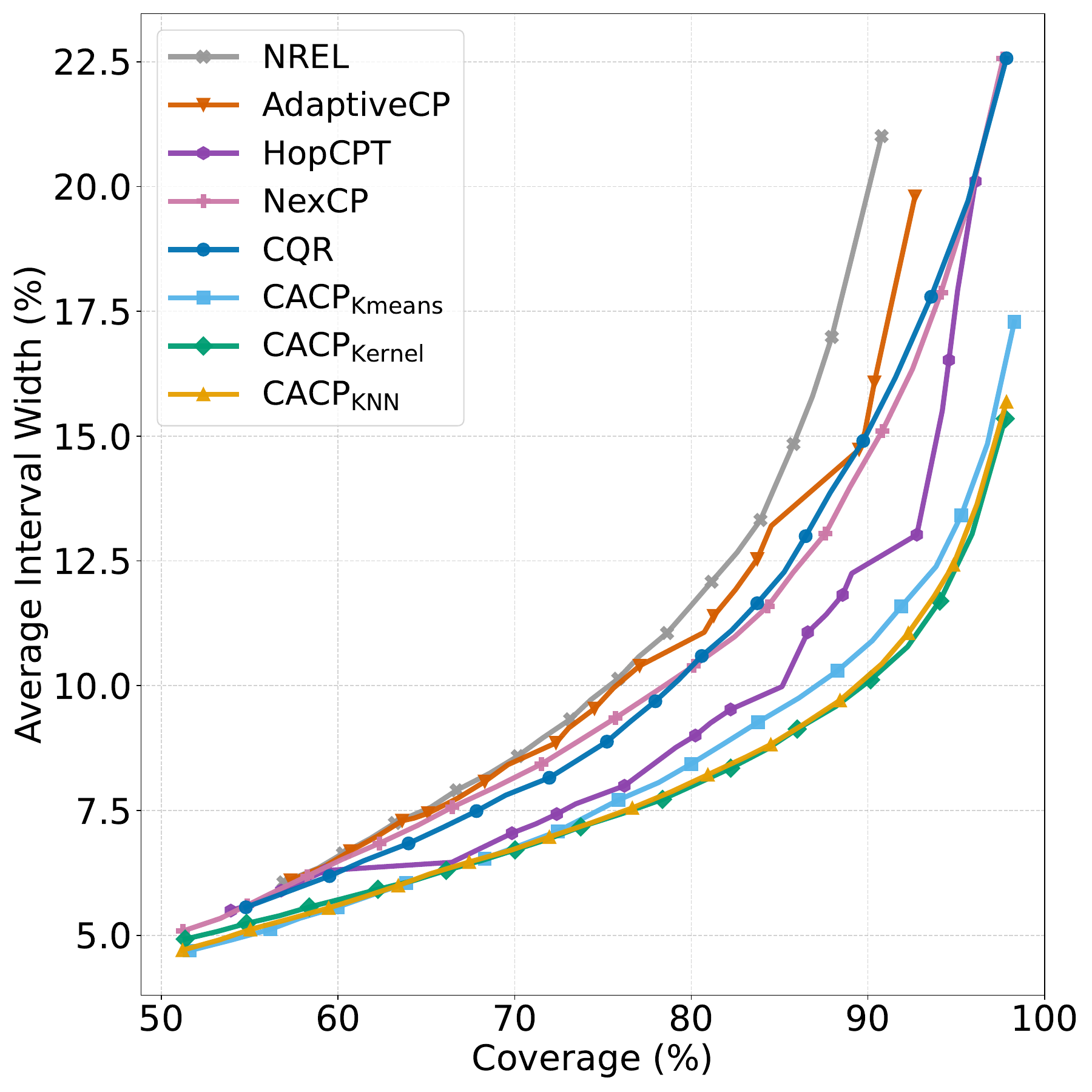}
        {\small (a) SPP system-level}
    \end{minipage}
    \hfill
    \begin{minipage}[t]{0.32\linewidth}
        \centering
        \includegraphics[width=\linewidth]{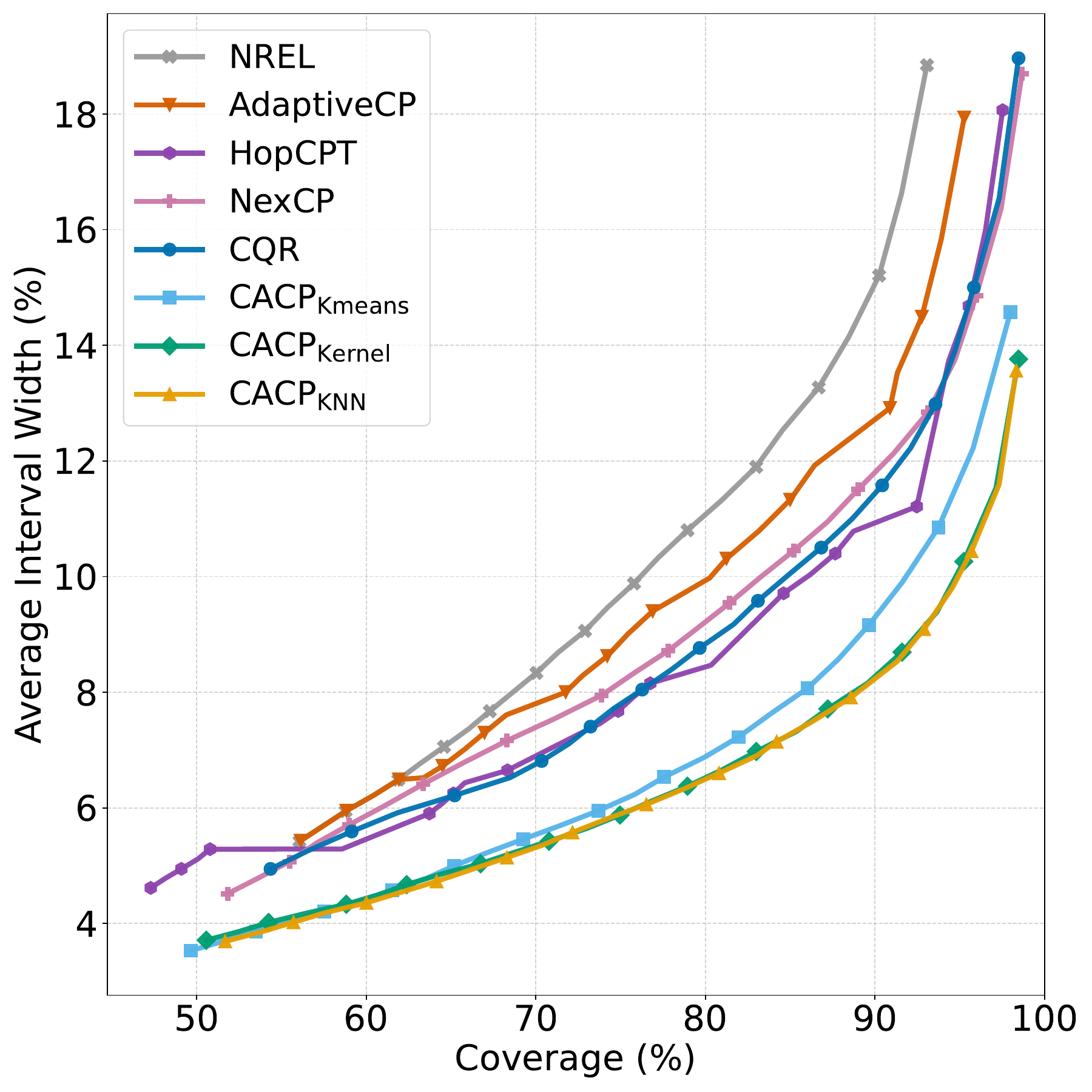}
        {\small (b) MISO system-level}
    \end{minipage}
    \hfill
    \begin{minipage}[t]{0.32\linewidth}
        \centering
        \includegraphics[width=\linewidth]{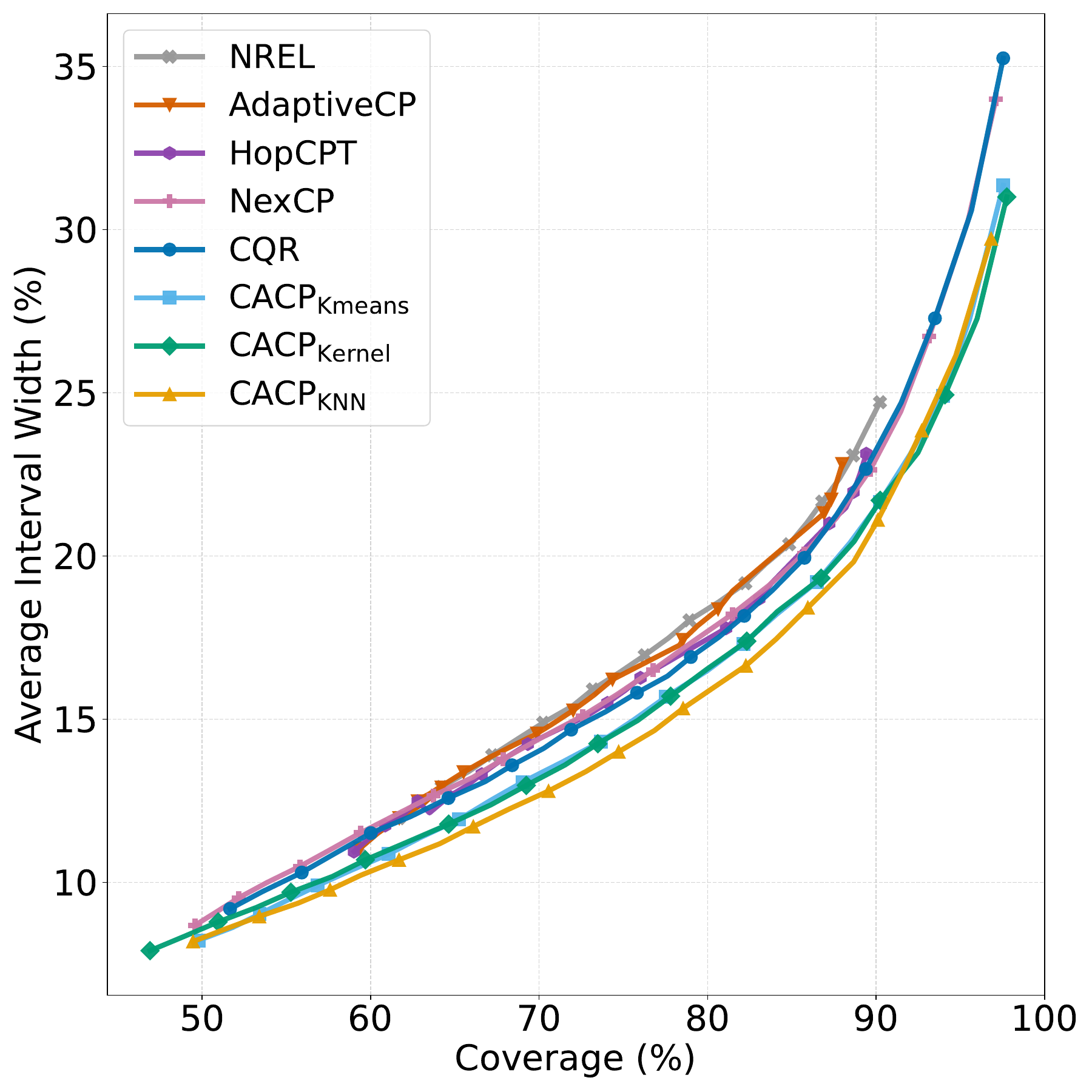}
        {\small (c) ERCOT system-level}
    \end{minipage}

    \caption{Coverage vs.\ AIW trade-off for different CP methods across system-level forecasts in three ISOs. \Method{} methods consistently achieve lower average interval widths compared with baselines at similar coverage rates.}
    \label{fig:picp_aiw_all}
\end{figure*}

\begin{figure*}[!t]
    \centering

    \begin{minipage}[t]{0.32\linewidth}
        \centering
        \includegraphics[width=\linewidth]{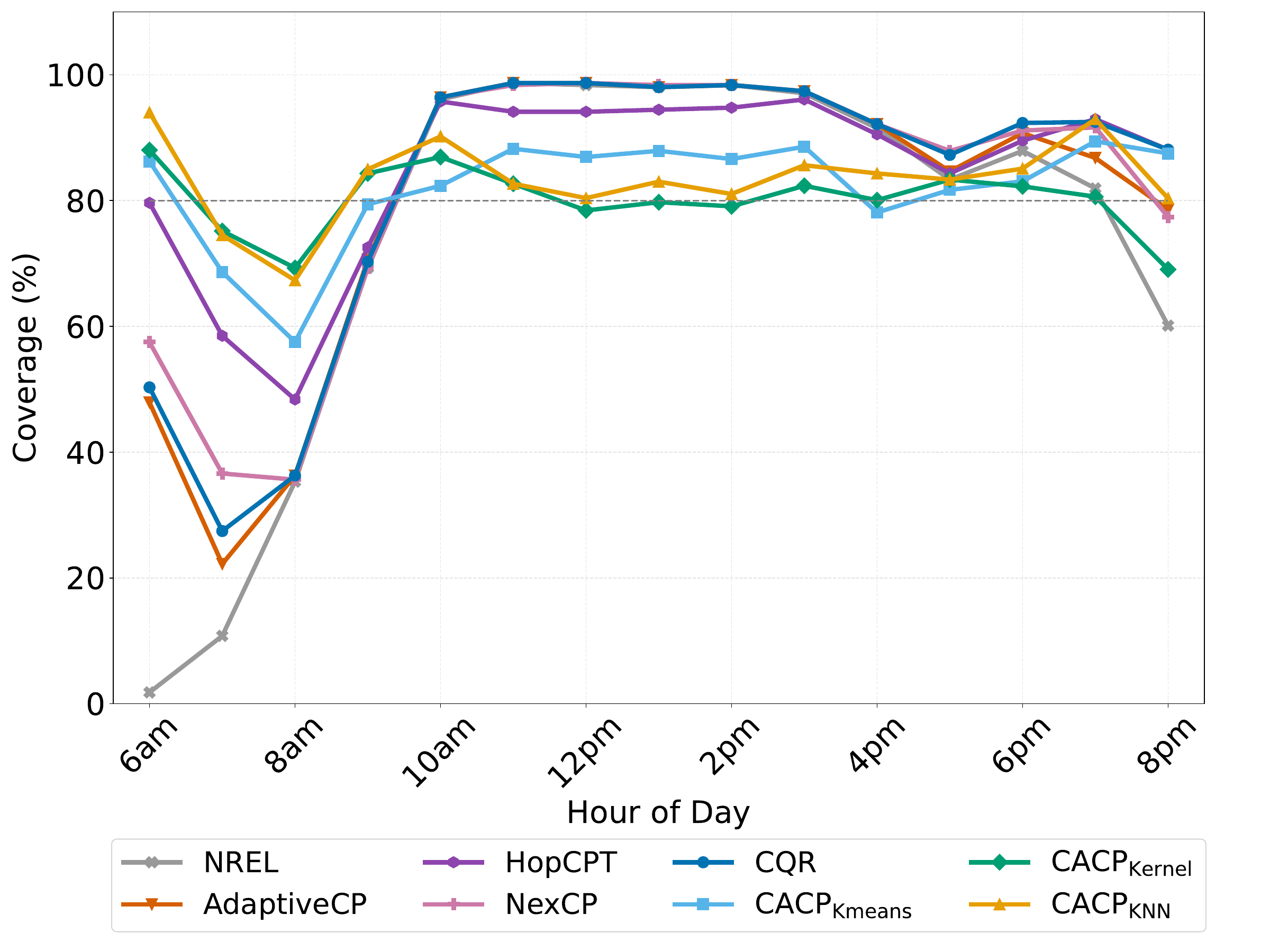}
        {\small (a) SPP system-level}
    \end{minipage}
    \hfill
    \begin{minipage}[t]{0.32\linewidth}
        \centering
        \includegraphics[width=\linewidth]{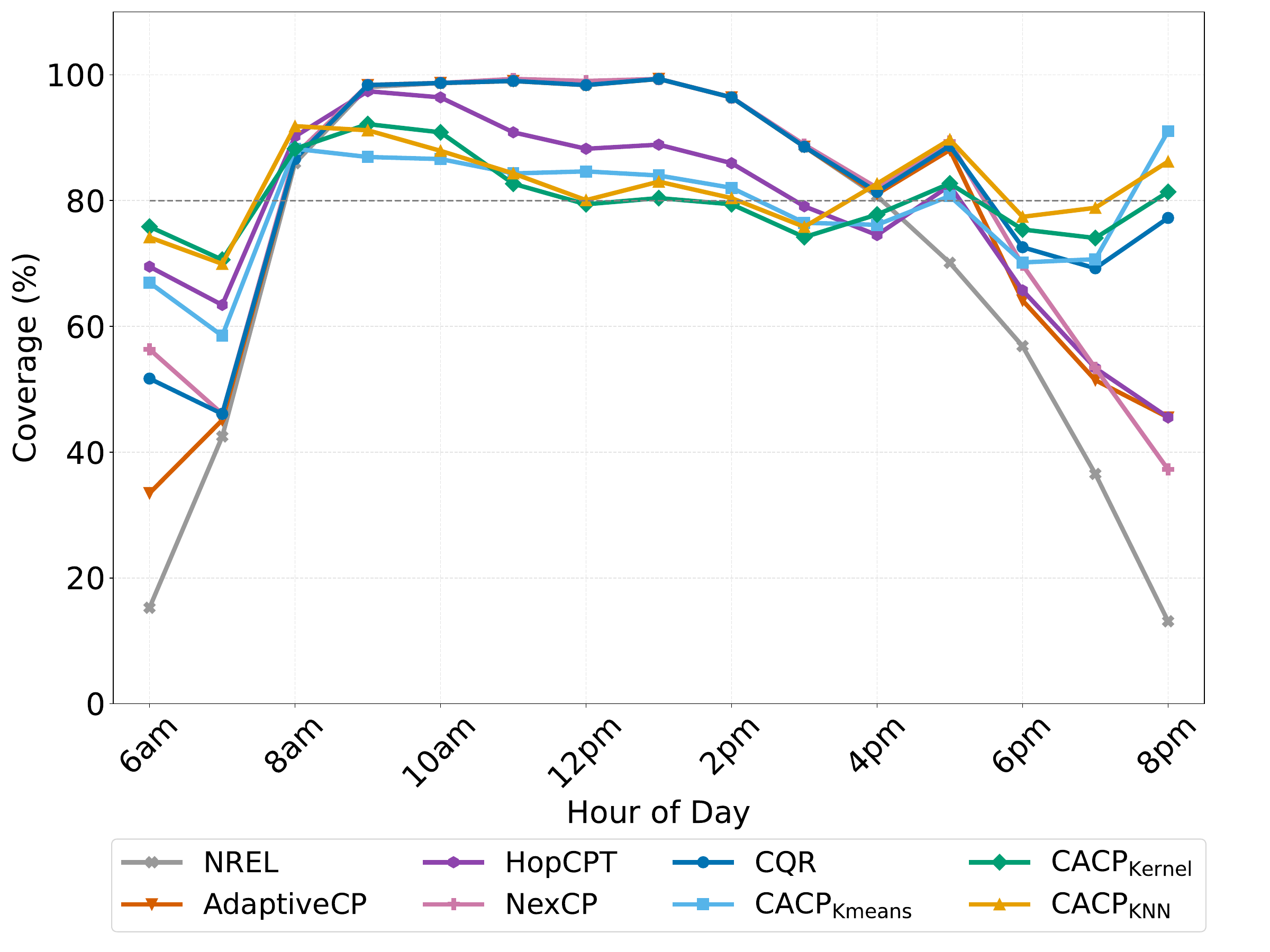}
        {\small (b) MISO system-level}
    \end{minipage}
    \hfill
    \begin{minipage}[t]{0.32\linewidth}
        \centering
        \includegraphics[width=\linewidth]{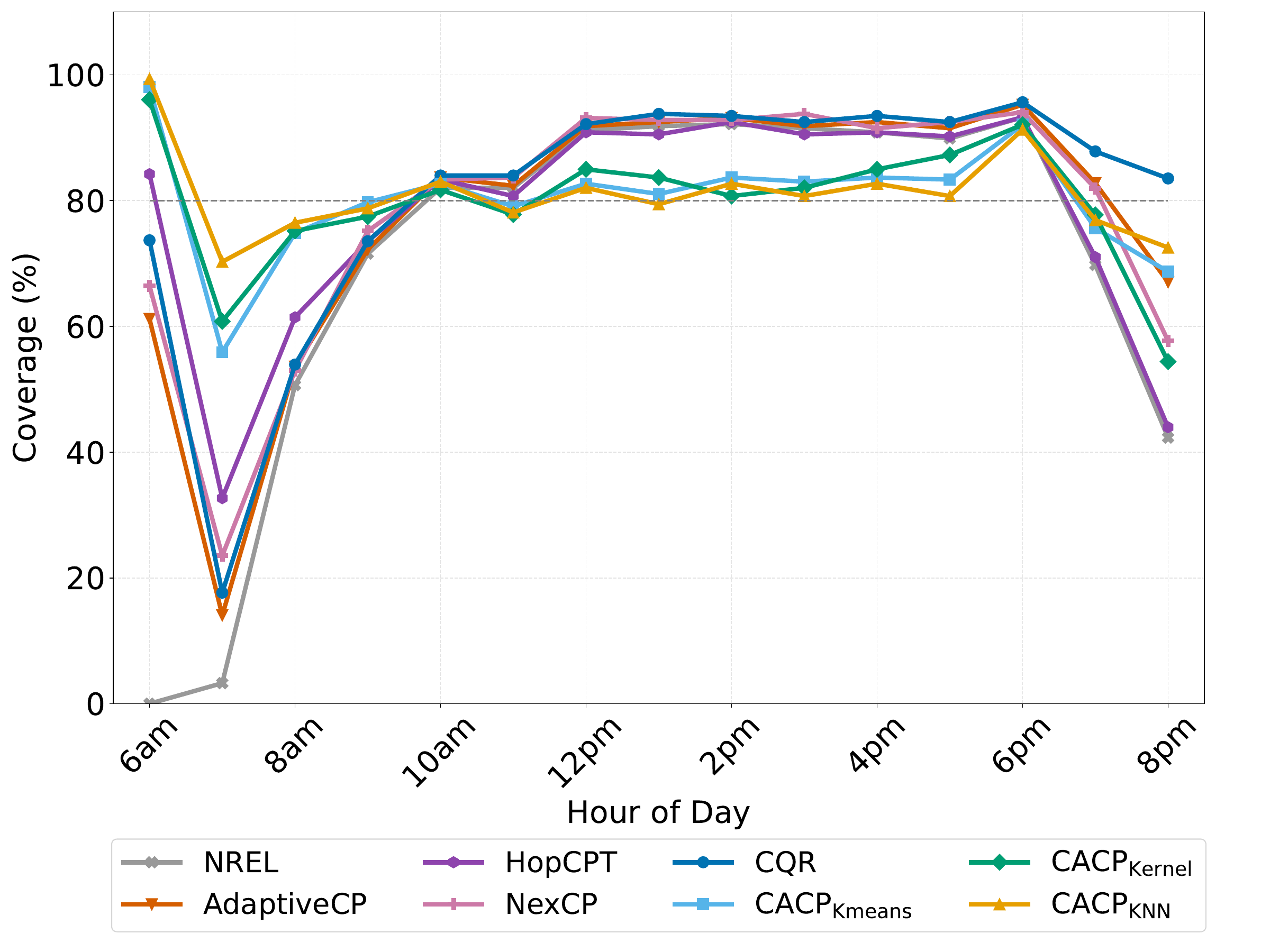}
        {\small (c) ERCOT system-level}
    \end{minipage}

    \caption{Conditional coverage for different hours of the day across three ISO system-level forecasts (SPP, MISO, and ERCOT) at target marginal coverage $80\%$. \Method{}-based methods maintain coverage close to the target across all hours, whereas baseline methods tend to under-cover during early morning and evening periods, and over-cover near midday peak hours.}
    \label{fig:hourly_picp_all}
\end{figure*}


\subsection{Variables Importance}
{\color{rev}
As discussed in Section~\ref{sec:cacp:features}, \Method{} leverages context-aware variables to retrieve calibration samples associated with operating regimes similar to the target prediction point. Since renewable energy systems exhibit strong temporal and meteorological nonstationarity, the relevance of different covariate groups changes over time. Consequently, \Method{} performs dynamic recalibration at each rolling prediction step, adaptively selecting the most informative variables for regime retrieval.

\begin{figure*}[!t]
    \centering
    \includegraphics[width=\linewidth]{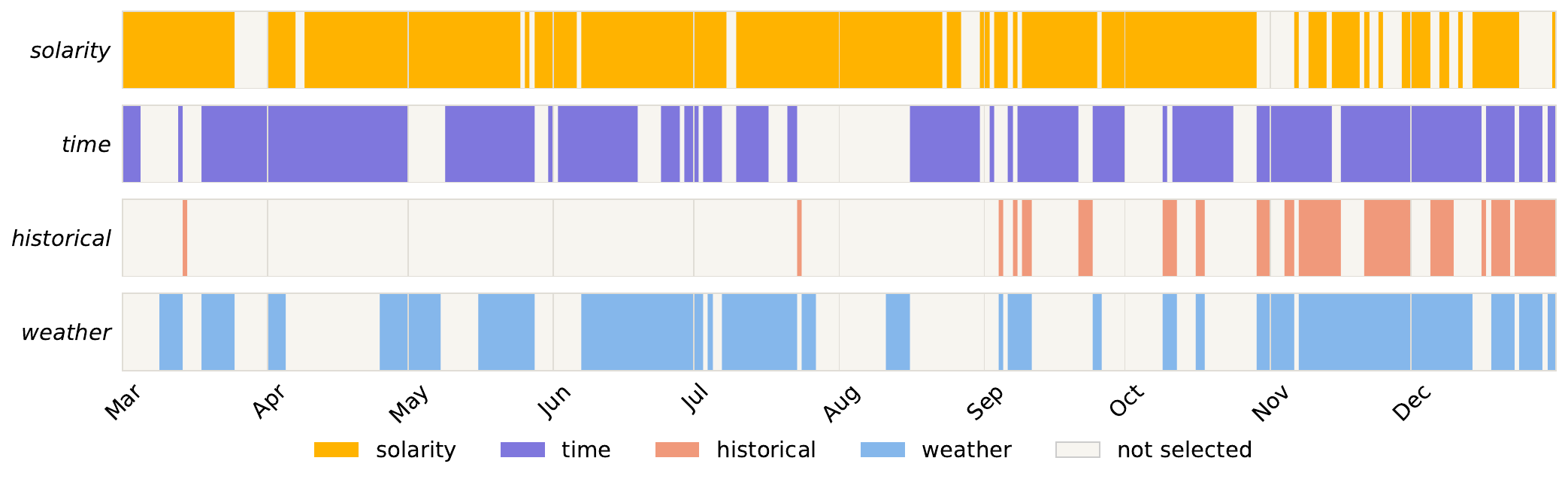}
    \caption{Temporal evolution of covariate-group selection for the \textsc{KNN} variant of \Method{} on the SPP system-level dataset. Colored segments indicate periods during which a covariate group was selected as informative for calibration, whereas gray regions denote periods where the group was not selected.}
    \label{fig:cov_spp}
\end{figure*}

\begin{figure*}[!t]
    \centering
    \includegraphics[width=\linewidth]{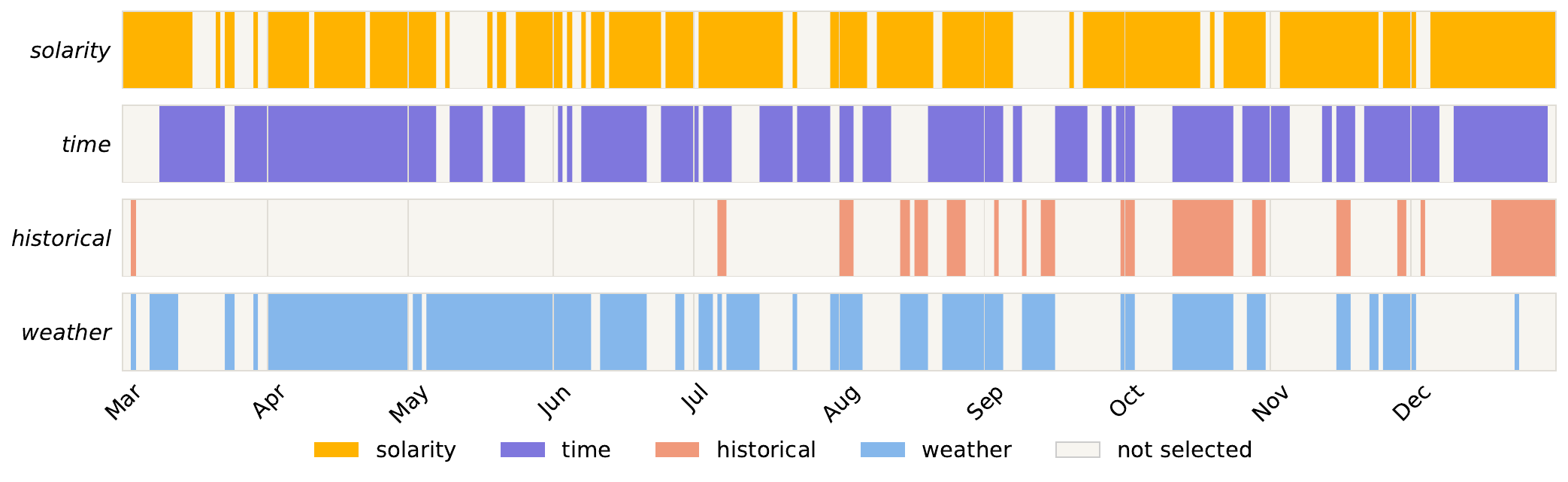}
    \caption{Temporal evolution of covariate-group selection for the \textsc{KNN} variant of \Method{} on the MISO system-level dataset. Colored segments indicate periods during which a covariate group was selected as informative for calibration, whereas gray regions denote periods where the group was not selected.}
    \label{fig:cov_miso}
\end{figure*}
}
Figures~\ref{fig:cov_spp} and~\ref{fig:cov_miso} illustrate the temporal evolution of covariate-group selection for the \textsc{KNN} variant of \Method{} on the SPP and MISO system-level datasets, respectively. Each column corresponds to a rolling recalibration window, while each row represents a covariate group used for context-aware weighting. Overall, both figures show that covariate importance is highly dynamic and regime-dependent rather than static throughout the forecasting horizon.

Several consistent patterns emerge across both systems. First, the \textit{Solarity} variables remain active during a large portion of the year, particularly throughout spring and summer months, highlighting the importance of daylight-cycle and solar-geometry information in defining forecast-error regimes. Second, the \textit{Time} embeddings exhibit strong but intermittent activation, suggesting that temporal periodicities become informative primarily during specific operational periods. Third, the \textit{Weather} variables become increasingly active during transition seasons and winter months, where atmospheric variability plays a larger role in renewable uncertainty.

At the same time, the figures also reveal meaningful differences between systems. Compared with SPP, the MISO system exhibits more persistent usage of weather-related variables throughout the year, suggesting a stronger dependence on meteorological conditions for identifying uncertainty regimes. In contrast, the SPP system relies more heavily on solarity-related variables during extended periods, indicating more stable solar-generation patterns driven by daylight dynamics. In both systems, the \textit{Historical} variables are selected less frequently and mainly during localized intervals, implying that short-term persistence alone is insufficient to characterize renewable forecast uncertainty across all operating conditions.

Overall, these results provide empirical evidence that different operating periods rely on different combinations of contextual information, supporting the central premise of \Method{} that renewable energy forecasting is governed by multiple latent operating regimes rather than a single stationary error process.

\section{Conclusion}
\label{sec:conclusion}
This paper introduced CACP, a context-aware CP framework for calibrating probabilistic forecasts in renewable energy systems. The proposed method was evaluated against multiple CP baselines for time series forecasting using large-scale datasets from MISO, SPP, and ERCOT, covering both system- and site-level solar generation forecasts provided by NREL. Experimental results showed that CACP achieves valid empirical coverage while producing sharper prediction intervals compared to existing baselines. In addition, CACP offers a favorable trade-off between predictive performance and computational efficiency, outperforming more resource-intensive methods such as HopCPT. The results highlight the effectiveness of the context-aware weighting mechanism in delivering more reliable conditional coverage. Specifically, the method reduces interval widths in regions where the base forecasts are overconfident and increases them during non-peak hours where underconfidence is more likely, adapting to local uncertainty patterns across time. 

Future work will extend this framework by incorporating additional covariates, such as weather data, and by applying it to wind power forecasting. 
Additional directions include exploring multi-dimensional and hierarchical settings, where context-aware weighting can be combined with the underlying spatio-temporal correlations to enhance forecast reliability.

\section{Acknowledgments}
\label{sec:ack}
This work was partially funded by NSF award 2112533 and by Los Alamos National Laboratory’s Directed Research and Development project, ``Artificial Intelligence for Mission" (ArtIMis).



%
\clearpage
\bibliographystyle{plain}  
\bibliography{ref}

\clearpage

\end{document}